\documentstyle[graphicx]{mn2e}

\newif\ifAMStwofonts

\title[The clustering of high redshift red galaxies in UKIDSS DXS SA22]{Clustering properties of high redshift red galaxies in SA22 from the UKIDSS DXS}
\author[Kim et al.]{J. -W. Kim$^{1}$\thanks{E-mail:
j.w.kim@durham.ac.uk} A. C. Edge$^{1}$, D. A. Wake$^{1,2}$ and J. P. Stott
$^{1,3}$\\
$^{1}$Institute for Computational Cosmology, Department of Physics, University 
of Durham, South Road, Durham DH1 3LE, UK\\
$^{2}$Department of Astronomy, Yale University, New Haven, CT 06520, USA\\
$^{3}$Astrophysics Research Institute, Liverpool John Moores University, Twelve 
Quays House, Egerton Wharf, Birkenhead, CH41 1LD}
\begin{document}

\date{Accepted ???? March ??. Received ???? March ??; in original form 5th March 2006}

\pagerange{\pageref{firstpage}--\pageref{lastpage}} \pubyear{2010}

\maketitle

\label{firstpage}

\begin{abstract}
Deep, wide, near-infrared imaging surveys provide an opportunity to study the clustering 
of various galaxy populations at high redshift on the largest physical scales. 
We have selected $1<z<2$ extremely red objects (EROs) and $1<z<3$ distant red 
galaxies (DRGs) in SA22 from the near-infrared photometric data of the UKIDSS Deep 
eXtragalactic Survey (DXS) and $gri$ optical data from 
CTIO covering 3.3~deg$^2$. This is the largest contiguous area studied
to sufficient depth to select these distant galaxies to date.
The angular two-point correlation functions and the real space 
correlation lengths of each population are measured and show that both
populations are strongly clustered and that the clustering cannot
be parameterised with a single power law. The 
correlation function of EROs shows a double power law with the inflection 
at $\sim$ 0.6$'$--1.2$'$ (0.6--1.2~h$^{-1}$~Mpc). The bright EROs ($K<18.8$) 
show stronger clustering on small scales but similar clustering on larger scales,
whereas redder EROs show stronger clustering on all scales.
Clustering differences
between EROs that are old passively evolved galaxies (OGs)
and dusty star-forming galaxies (DGs), on the basis of their $J-K$
colour, are also investigated. 
The clustering of $r-K$ EROs are 
compared with that of $i-K$ EROs and the differences are consistent
with their expected redshift distributions. 
The correlation function of DRGs is also well described by a double power law
and consistent with previous studies once the effects of
the broader redshift distribution our selection of DRGs returns are
taken into account.
We also perform the same analysis on smaller sub-fields to investigate the 
impact of cosmic variance on the derived clustering properties. Currently 
this study is the most representative measurement of the clustering of massive 
galaxies at $z>1$ on large scales.
\end{abstract}

\begin{keywords}
surveys-galaxies: evolution - galaxies: photometry - cosmology: observations - infrared: galaxies.
\end{keywords}

\section{Introduction}

The Lambda cold dark matter ($\Lambda$CDM) paradigm predicts that small scale 
structure has developed by accretion and mergers within the large scale structure
generated by primordial mass fluctuations.
In addition, the galaxies tracing this structure
are themselves embedded and have evolved in the dark 
matter halos (White \& Frenk 1991). The halo properties such as 
abundance, distribution and density profile depend on the mass of halo 
(Cooray \& Sheth 2002). As a result, the formation and evolution of galaxies 
is affected by the halo 
mass (Eke et al. 2004; Baugh 2006). Therefore the clustering properties of 
galaxies can be related to the distribution of dark matter halos, and hence offer 
an important insight into the relationship between the halo and the galaxies
within it. For instance, Wake et al. (2008) 
used correlation functions and halo models to demonstrate that Luminous Red 
Galaxies (LRGs) are frequently located in the centre of the most massive
dark matter haloes, and that changes in their small scale clustering
with redshifts can constraint LRG-LRG merger rates.
At higher redshifts, Mo \& White (2002) pointed out that halos of given mass are 
expected to be more clustered from simulations. 
Observationally Foucaud et al. (2010) have demonstrated that galaxies 
with higher stellar masses are more clustered, and galaxies with a fixed stellar 
mass are also more clustered at higher redshift. Hartley et al. (2010) have
also shown that passive, red galaxies are more clustered than
star-forming, blue galaxies of similar absolute magnitudes at 0.5$<z<$3.0.
Quadri et al. (2008) found that a double power law was required to describe 
the correlation function of DRGs at $2 < z < 3$, but were unable to fit their 
clustering measurement and space density simultaneously
using the halo model. Tinker, Wechsler \& Zheng (2010) showed that using a more realistic 
halo model they could better fit this clustering measurement, but 
they still required that the observed region be more
clustered part of the universe than is typical.
However, most 
observational studies of clustering and the halo model have concentrated 
on relatively low redshifts ($z<1$) so their evolution has been 
poorly constrained.  With the advent of large
near-infrared surveys, it is now possible to apply these techniques
to more distant galaxies.
The study of the angular clustering of $z>1$ galaxies is particularly
powerful as the near constant angular diameter distance in the 
$1<z<3$ range means that angle and comoving distance are much more
closely linked than at lower redshift. Therefore, any characteristic
distance (halo transition or sound horizon at last recombination) will
translate to a small range in angle.

Many colour criteria have been used to select 
 high redshift galaxies. One of the crudest but most direct methods is
to select galaxies with a large colour difference 
between optical and infrared wavelengths ($I-K>4$) (Elston et al. 1988)
to select Extremely Red Objects (EROs). This technique
preferentially selects the most massive galaxies at 
$z>1$ which tend to be dominated by an evolved stellar population
with a pronounced 4000$\rm \AA$ break. However, the spectra of 
EROs show evidence for a significant population of dusty, star-forming 
galaxies (Pozzetti \& Mannucci 2000; Smail et al. 2002; Roche et al. 2002; 
Cimatti et al. 2002, 2003; Moustakas et al. 2004; Sawicki et al. 2005; 
Simpson et al. 2006; Conselice et al. 2008; Kong et al. 2009).
EROs are strongly clustered (Daddi et al. 2000; Roche et al. 2002, 2003; Brown 
et al. 2005; Kong et al. 2006, 2009), and reside in massive dark matter halos 
(Gonzalez-Perez et al. 2009).

Alternatively, a large near-IR colour difference, $J-K>2.3$, can be
used to select Distant Red Galaxy (DRGs) (Franx et al. 2003) that
are predominantly at $z>2$. As with EROs, this simple colour cut
selects galaxies with a range of redshift (Lane et al. 2007; Quadri 
et al. 2007) and both dusty star-forming and
passive galaxies (Labb\'{e} et al. 2005; Papovich et al. 2006).
DRGs are also highly clustered (Grazian et al. 2006; Foucaud 
et al. 2007). 
Perhaps the most striking evidence for
clustering of high redshift galaxies is the angular clustering
of {\it Herschel} SPIRE sources presented in Cooray et al. (2010) 
where the large scale clustering and its distinct curvature are
unambiguously detected for $z>1$ FIR selected galaxies that
are presumably the progenitors of present day massive, early-type galaxies
from the similarity of their clustering properties.

The addition of a third photometric datapoint can improve the discrimination
of high redshift galaxy selection. For instance, for galaxies with a
strong UV continuum
 Steidel \& Hamilton (1992) used an optical colour criteria 
to identify Lyman Break Galaxies (LBGs) at $z>3$ and Adelberger et al. (2004) used the
same filter system but selected galaxies at $z\sim$1.7 and 2.3, termed BM/BX galaxies. 
The broadest photometric selection method was suggested by  Daddi et al. (2004)
using the B, z and K filters to identify star-forming (sBzK) and passively evolved 
(pBzK) galaxies at $z>1.4$. 

With all of these photometric selection methods,
the small field of view of imaging cameras has hampered 
the accurate measurement of large scale clustering. 
In particular, the lack of wide field near-IR instruments has prevented the
detection of distant, passive galaxies since the bulk of their emission is 
redshifted to longer wavelengths. However, new wide and deep near-IR surveys now
provide an opportunity to investigate the clustering properties of galaxies 
at high redshift. In this paper, we use the wide contiguous near-IR data from 
5th Data Release (DR5) of the 
 Deep eXtragalactic Survey Data (DXS),  which is the sub-survey of 
UK Infrared Telescope Infrared Deep Sky Survey (UKIDSS) (Lawrence et al. 2007), 
in conjunction with $gri$ 
optical data from the CTIO 4m to measure the clustering properties of EROs and DRGs, and 
discuss the clustering properties with various criteria.

In Section 2, we describe the compilation of UKIDSS DXS and optical data
for the SA22 field. Then 
the analysis method for near-IR and optical data and determining clustering 
properties are described in Section 3. The results are presented in Section 4, 
and discussed in Section 5. Unless otherwise 
noted, the photometry is quoted in the Vega scale. We also assume $\Omega_{m}=$0.27, 
$\Lambda=$ 0.73 and $H_{o}=$ 71 km s$^{-1}$ Mpc$^{-1}$.

\section{Observation}

\subsection{UKIDSS DXS}

The UKIRT Infrared Deep Sky Survey began in 2005 and consists of 5 sub-surveys covering 
various areas and depths (Lawrence et al. 2007). The UKIDSS uses the 
Wide Field Camera (WFCAM, Casali et al. 2007) mounted on the UK Infrared 
Telescope (UKIRT). Of the 5 sub-surveys, the DXS is a deep, wide survey 
mapping 35 deg$^2$ with 5$\sigma$ point-source sensitivity of $J\sim22.3$ 
and $K\sim20.8$. It is comprised of 4 fields and aims to create photometric 
samples at $z\sim1-2$. 

WFCAM is composed of four Rockwell Hawaii-II 2K$\times$2K array 
detectors (Casali et al. 2007). The pixel scale is 0.4 arcsec/pixel, so the 
size of each detector is 13.7$\times$13.7 arcmin$^2$. The relatively 
large pixel scale can lead to an undersampled point spread function (PSF). To 
avoid this problem microstepping is applied. In addition there are gaps 
between detectors, and the width is similar with the size of a detector. 
Therefore  4 exposures are needed to make a contiguous image, i.e. 
4$\times$4 image tiling 0.8 deg$^2$.

In this study we deal with $\sim$3.3 deg$^2$ SA22 field centred on $\alpha=$ 
22$^{h}$ 19$^{m}$ 17.0$^{s}$, $\delta=$ +00$^{d}$ 44$^{m}$ 00.0$^{s}$ (J2000) 
from UKIDSS DR5. Since it is composed with four 0.8 deg$^2$ 
fields from DXS SA22 1 to 4, there is a 16-point mosaic to cover the field. 
The DXS SA22 1 corresponds to the south-west part of our field, and SA22 2, 3 and 4 
are south-east, north-east and north-west, respectively. Seeing conditions of 
all images were $\sim$0.9$''$ at $J$ and $\sim$0.8$''$ at $K$.

\subsection{Optical data}

Optical $gri$ images were obtained from 4m Blanco Telescope in Cerro Tololo 
Inter-American Observatory (CTIO) in September 2006 as complement to
the DXS as the SA22 field lacked any wide field optical imaging.
The observations were performed by 
Mosaic II CCD composed with eight 2K$\times$4K detectors. Each exposure 
covers 36$\times$36 arcmin$^2$, and 9 fields were observed to map $\sim$3.3 
deg$^2$ UKIDSS DXS 
SA22 field. Total exposure times are 1,800 seconds for $g$ and 3,000 seconds 
for $r$ and 5,400 seconds for $i$. In addition, 5- or 9-point dithering 
methods were applied for $g$, $r$ and $i$ 
respectively. The seeing was 1-2$''$ for $g$ and 1-1.5$''$ for $r$ and $i$.

\section{Analysis methods}

\subsection{UKIDSS DXS}

The UKIDSS standard pipeline creates a catalogue by merging detected objects 
for each array. However, Foucaud et al. (2007) pointed out that creating a 
contiguous image before extracting the catalogue is more helpful to optimise 
depths in overlap regions and to make a homogeneous image.  
There are also some known issues that require particular care with
analysing WFCAM data. Dye et al. (2006) discussed 
various artefacts of UKIDSS Early Data Release (EDR), including
 internal reflections and electronic cross-talk. The former have been 
eradicated by additional baffling and restricting the observations
with respect to Moon angle but the latter still have to be removed. Cross-talk
is straightforward as each cross-talk feature is a fixed number
of pixels from a bright star and can hence be flagged (see below).
The standard CASU detection pipeline is not 
optimised for galaxy photometry so we choose to create
our own photometric catalogues from mosaicked images.

Firstly, four contiguous images were created from the reduced images 
including astrometric and photometric information from UKIDSS standard 
pipeline. As mentioned above, we used $\sim$ 3.3 deg$^2$ field 
which is composed of four 0.8 deg$^2$ fields. Instead of creating one large 
contiguous image, four separate images were made, since the 
integration time for each exposure was changed from 5 seconds to 10 seconds
after the first UKIDSS observing period. 
Before stacking, the 
flux scale of each array was calculated by using the {\it zeromag} from the UKIDSS standard 
pipeline. The SWarp software, which is a resampling and stacking 
tool (Bertin et al. 2002), was run to stack the images. A weight map was created 
by SWarp, and bad regions at the detector edges or around saturated stars were masked 
by visual inspection.

Secondly, objects were extracted from these contiguous images by running 
SExtractor (Bertin \& Arnouts 1996). 
To measure colours of objects, the flux has to be measured for the same part 
of each object. Therefore the $J$-band image was transformed into the 
$K$-band image frame 
by IRAF task GEOMAP and GEOTRAN which calculate a transforming equation and 
transforms the images, respectively.  
Also, in order to minimise the fraction of spurious objects, 
various combinations of thresholds and detection area were tested.
Finally, the $K$-band image was used as a detection image 
to detect faint objects in the other bands. Colours were calculated 
using 2-arcsec aperture 
magnitudes for each band, and total magnitudes in $K$ were measured using the
SExtractor AUTO 
magnitude. A photometric calibration was applied by using the calibrated 
aperture magnitude of the point source catalogue from UKIDSS standard pipeline 
which was calibrated from the  Two Micron All Sky Survey (2MASS, Skrutskie et 
al. 2006).

Thirdly, spurious objects such as cross-talk images and diffraction spikes 
were removed. The cross-talk images were located at multiples of 256 pixels 
from bright objects (Dye et al. 2006). We selected cross-talk candidates 
related to bright stars ($K<16$) by position. All potential cross-talk images from
the brightest stars ($K<13$) were removed. However, those from  fainter 
stars ($13<K<16$) contain a fraction of real objects. Fortunately, since cross-talk 
images of fainter hosts accompany dark spots around them, these spots were 
used to determine whether candidates are cross-talk images or real objects. 
In addition, spurious objects detected on diffraction spikes were selected 
by position from bright stars
and removed. In total, 6 per cent of the objects were removed as spurious objects 
using these methods. We detected 303,473 bona fide objects from the masked 3.07 deg$^2$ field.

Finally, a completeness test was performed to check the photometric quality.
We randomly added 1,000 artificial stars into each image, and ran SExtractor 
again with the same parameters. This process was repeated 10 times with different 
artificial stars. Figure 1 shows the completeness of each field for $J$ (top) 
and $K$ (bottom). Horizontal and vertical lines indicate the 90 per cent 
completeness and the design depth of DXS, respectively. The completeness 
of DXS SA22 4 at $J$ is significantly lower than others. This was caused by 
the relatively high background value in this field. However, the other fields successfully 
 reach the target magnitude limit in both bands.

% figure : completeness test
\begin{figure}
\includegraphics[width=8cm]{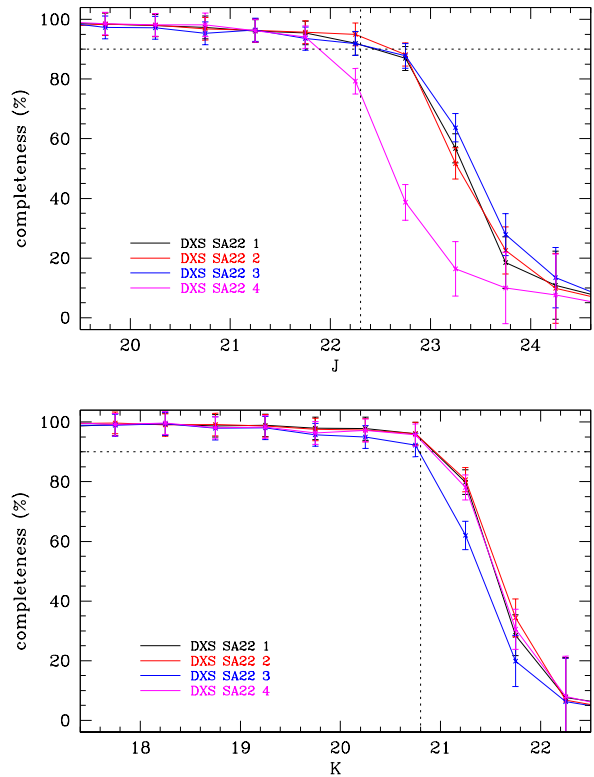}
\caption{Completeness test result. Horizontal and vertical lines indicate 
90 per cent completeness and the target limiting magnitude of the DXS, respectively.}
\end{figure}

\subsection{Optical data}

We followed the standard image reduction sequence for mosaic CCD, namely
bias subtraction, flat-fielding by dome and sky flat images, masking bad 
pixels and removing cross-talk artefacts and cosmic rays. 
The USNO-A2 catalogue was used to improve the astrometric solution for each field.
This solution was also applied to project images so they had the same scale
and astrometry using the IRAF tasks GEOMAP and GEOTRAN.
Finally, these projected images were 
combined by using median values, and exposure maps were used as a weight map.

As with the near-IR data, colours of galaxies have to be measured from 
the same part of each object. Therefore, images for the same field have to be
matched to have the same seeing. To do this, the
better seeing images were degraded using the PSFMATCH task in IRAF. The DAOPHOT package 
was used to select unsaturated stars and to create PSF kernels.
These kernels were then used to degrade better seeing images to the worst seeing.

In order to detect objects and measure fluxes, the same strategy used for the 
near-IR data was applied. SExtractor (Bertin \& Arnouts 1996) was run in 
dual mode. The $i$-band image was used as the detection image. Saturated 
stars and their halos were masked in the weight image to remove unreliable 
objects. Various threshold values were tested, and the value minimising 
spurious objects was selected. Finally, since the seeing of the CTIO 
imaging was worse than the DXS, 
ISO and AUTO magnitude were used to estimate colours instead of aperture 
magnitude and total magnitude.

Photometric calibration was performed using the Sloan Digital Sky Survey (SDSS, 
York et al. 2000) catalogue. Aperture colours from SExtractor were 
calibrated to those of SDSS, and the absolute flux calibration was determined using the
 total magnitudes in $i$ with respect to those in the SDSS. 
Finally, we removed unreliable photometric results using
a magnitude cut. To remove saturated stars, objects brighter than $i<17.6$ 
were replaced by the equivalent SDSS data. In addition, since the number density of  
objects decreases sharply at $i>24.6$ and the completeness computed from 
artificial stars is $\sim$85 per cent at $i=24.6$, we extracted only
objects brighter than $i=24.6$. A total of 302,402 objects were extracted 
for the masked 2.45 deg$^{2}$ optical catalogue.

\subsection{Matching optical and near-IR catalogues}

% figure : CCD comparison
\begin{figure}
\includegraphics[width=8cm]{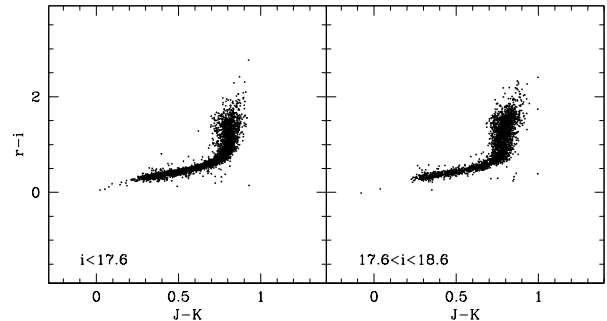}
\caption{The $r-i$ vs. $J-K$ two-colour diagram for point sources of 
$i<17.6$ (SDSS source) and $17.6<i<18.6$ (CTIO source).}
\end{figure}

To create the final catalogue, optical and near-IR catalogues were combined. 
Firstly, the near-IR catalogue was matched with the optical catalogue 
with a 1 arcsec distance, and the
average offsets were measured. The offsets were applied to the near-IR catalogue 
and then the offsets were recalculated. This process was repeated until the 
average offset was less than 0.1 arcsec. The calculated offsets were 0.05 
arcsec toward the west and 0.43 arcsec toward the north. Finally, the calculated 
offsets were removed from the near-IR catalogue, and the closest 
optically detected object within 1 arcsec was taken as the counterpart. 

A Galactic extinction  correction was applied to  all objects. The coordinate of 
each object was used to measure the reddening value from a dust map (Schlegel, 
Finkbeiner \& Davis 1998). Then the values were transformed into extinction 
values for each band using the coefficients in Schlegel et al. (1998).

Due to the seeing differences, different aperture magnitudes were used for the optical and near-IR 
catalogues. To ensure this didn't affect our optical to near-IR colours 
we compared the two-colour diagram of point sources. The 
point sources were selected by the magnitude difference measured between 0.4$''$ and 
2$''$ diameter apertures in $K$. Figure 2 shows the two-colour diagram of 
point sources for $i<17.6$ (left) and $17.6<i<18.6$ (right); there is no significant
difference. We also note that there are no appreciable field-to-field variations. Since we 
combined $i<17.6$ SDSS sources, figure 2 indicates that our colour is not 
affected by the different methods used to  measure fluxes.

\subsection{Angular correlation function and correlation length}

% figure : IC
\begin{figure}
\includegraphics[width=8cm]{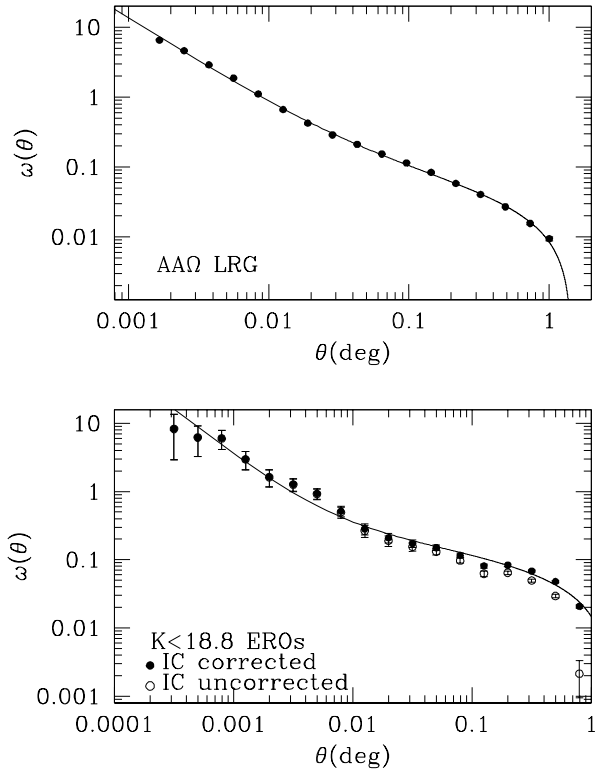}
\caption{Fitted results by the assumed correlation function for AA$\Omega$ 
LRG (top) in Sawangwit et al. (2010) and our $K<18.8$ EROs (bottom).
Open and filled circles in the bottom panel show the correlation function before and after 
correcting for the integral constraint.}
\end{figure}

The angular two-point correlation function is the probability of finding 
galaxy pairs with respect to a random distribution (Peebles 1980). Usually the 
estimator from Landy \& Szalay (1993) is used to estimate the angular 
two-point correlation function:
\begin{equation}
\omega_{obs}(\theta)=\frac{DD(\theta)-2DR(\theta)+RR(\theta)}{RR(\theta)},
\end{equation}
where DD is the number of observed data pairs with separation interval 
$[\theta,\theta+\Delta\theta]$.  For this study we used $\Delta log\theta=$0.2.
DR and RR are respectively the mean data-random and random-random pairs in the same interval.
All  pair counts are  normalised to have same totals.

In order to count DR and RR, we generated the random catalogue with 100 times 
more random points than the observed data sample. Our random catalogue covered
exactly the same angular mask as our data, including the 
exclusion of regions around bright stars.

The error of each point in the correlation function was estimated 
from the poissonian variance of the DD pairs, 
\begin{equation}
\delta\omega_{obs}(\theta)=\frac{1+\omega_{obs}(\theta)}{\sqrt{DD}}
\end{equation} 
where DD is the unnormalised value.

The restricted area of our observations leads to the negative offset of the
observed angular correlation function which is known as the integral 
constraint. Therefore, the relation between the real correlation function 
($\omega(\theta)$) and the observed correlation function 
($\omega_{obs}(\theta)$) is 
\begin{equation}
\omega_{obs}(\theta)=\omega(\theta)-\sigma^{2},
\end{equation}
where $\sigma^{2}$ is the integral constraint (Groth \& Peebles 1977). 

If $\omega(\theta)$ is known, the integral constraint can be 
calculated numerically from the equation in Roche et al. (1999), 
\begin{equation}
\sigma^{2}=\frac{\sum{RR(\theta)\omega(\theta)}}{\sum{RR(\theta)}}.
\end{equation}

In most previous studies, $\omega(\theta)=A_{\omega}\theta^{-\delta}$ was 
assumed for the correlation function with a slope fixed at $\delta=$0.8.
However, applying a single power law is not appropriate for the correlation 
function of our samples over the range of angle achieved in this study 
(see section 4). Also even using a double power law can lead to an uncertain integral 
constraint value if the slope fitted to the larger scales is shallow. This
can lead to a greatly inflated integral constraint on scales larger
than the sound horizon at recombination ($\sim 100 h^{-1}$Mpc) beyond which the
angular clustering should be negligible but is predicted to be large.
This is particularly important for this study as the largest scales sampled
here are comparable to the natural cut-off in clustering that has been
demonstrated directly from larger scale surveys by, for example, Maddox et al. (1990)
and Sawangwit et al. (2010). To avoid this overestimation of the integral
constraint we assume the correlation function has a form of $\omega({\theta})=\alpha_{1}\theta^{-\beta_{1}}+
\alpha_{2}\theta^{-\beta_{2}}-C$, where $C$ is a constant. This functional form 
provides a good fit to the angular correlation function of AA$\Omega$ LRGs 
in Sawangwit et al. (2010) as shown in the upper panel of their figure 3.
With this assumed functional form, we calculated the integral constraints of our samples by an iterative 
technique with equation (1), (2), and (3). The bottom panel in figure 3 shows the 
example of $K<18.8$ EROs before and after correcting for the integral 
constraint (open and filled circles) with fitted result (solid line). It is also 
confirmed that the assumed form fits our results well. After correcting for 
the integral constraint with the assumed form, we used the simple power law, 
$\omega(\theta)=A_{\omega}\theta^{-\delta}$, to measure amplitudes and 
slopes of each sample on small and large scale (see section 4 and 5 for details).

The observed angular correlation function corresponds to a projection of
the real space correlation function, which is assumed to have a power law form.
\begin{equation} 
\xi(r)=\left(\frac{r}{r_0}\right)^{-\gamma}
\end{equation}
where $\gamma=1+\delta$. The value of $r_{0}$, the correlation length,
can be calculated by Limber's transformation from the amplitude of angular 
two-point correlation function (Limber 1953; Peebles 1980). In this study, we 
used the relation in Kova{\v c} et al. (2007). The relation is as follows:
\begin{equation}
A_{\omega}= r_0^{\gamma}\sqrt{\pi} \frac{\Gamma(\frac{\gamma-1}{2})}{\Gamma(\frac{\gamma}{2})}\frac{\int_{0}^{\infty} F(z) D_A^{1-\gamma}(z) N_{corr}(z)^2 g(z) dz}{\biggl[\int_{0}^{\infty} N_{corr}(z) dz \biggr]^2}
\end{equation}
where $A_{\omega}$ is the amplitude of angular correlation function, $\Gamma$ 
is the gamma function, $D_{A}(z)$ is angular diameter distance 
calculated by the Javascript Cosmology Calculator (Wright 2006) and 
$N_{corr}(z)$ is the redshift distribution of the detected galaxies. 
In addition, $g(z)$ is given by 
\begin{equation}
g(z)=\frac{H_o}{c} \biggl[ \left(1+z\right)^2 \sqrt{1+ \Omega_M z +\Omega_{\Lambda} \left[\left(1+z\right)^{-2}-1\right]} \biggr]
\end{equation}
for standard cosmological parameters and $F(z)$ is a redshift evolution term. 
Blanc et al. (2008) point out $F(z)=(1+z)^{-(3+\epsilon)}$, where values of 
$\epsilon=$-1.2 corresponds to fixed clustering in comoving coordinates, $\epsilon=$0.0
corresponds to fixed clustering in proper coordinates and
and $\epsilon=$0.8  is the prediction from linear 
theory, Brainerd, Smail and Mould (1995). In this study, we 
assume the first case, that the clustering is fixed in comoving 
coordinates. In addition, we use a power law slope, $\delta$, determined from
$i-K>4.5$ and $K<18.8$ EROs to calculate the correlation lengths of various 
EROs.

We generate the redshift distribution for each sample using the photometric 
redshifts produced by the NEWFIRM Medium Band Survey 
(NMBS; van Dokkum et al. 2009; Brammer et al. 2009; van Dokkum et al. 2010). 
This survey images two 0.25 deg$^2$ areas in the AEGIS (Davis et al. 2007) and 
COSMOS (Scoville et al. 2007) fields in 5 medium band filters in the 
wavelength range 1-1.7$\mu$m as well as the standard $K$-band. The addition of 
these 5 medium band near-IR filters to the already existing deep multi-band 
optical (CFHTLS) and mid-IR (Spitzer IRAC and MIPS) enables precise photometric 
redshifts ($\sigma_{z}/(1+z)<$ 0.02) to be determined for the first time for 
galaxies at z $>$ 1.4, where the main spectral features are shifted into the 
near-IR. Although the NMBS can miss the rarest, bright galaxies because of the 
small surveyed area, NMBS imaging is significantly deeper than the DXS so we 
are able 
to directly apply all the same selection criteria that we apply to each sample 
in this paper in order to determine a meaningful redshift distribution. We make use of 
the full photometric redshift probability distribution functions (PDF) output 
by the EAZY photometric redshift code (Brammer, van Dokkum \& Coppi 2008) 
that has been used to produce the NMBS photometric redshift catalogue. For 
each sample, our redshift distribution is defined as the sum of all the PDFs 
for the galaxies passing the appropriate colour and magnitude selection cuts. 
The redshift distributions of EROs show different trends with various 
selection cuts in magnitude and colour.
On the one hand, magnitude limited EROs are predominantly at 
$1<z<2$ with a significant peak at 
$z\sim$1.2, and a tail to higher redshift that is most apparent for fainter EROs.
On the other hand, colour limited EROs show a much broader redshift distribution
where the mean increases at higher values of $i-K$.
Using the bluest cut of  $i-K>3.96$, a significant population ($>20$\%) of 
$z<1$ objects is included.
For DRGs, the brightest ($K<18.8$) are concentrated at $z\sim$1.1 and
the faintest ($18.8<K<19.7$) are more broadly distributed between
$1.3<z<1.9$.

In order to estimate the uncertainty in the correlation length, a Monte Carlo 
approach was applied. First 1,000 amplitudes having a normal distribution 
were generated with the error in amplitude. Then correlation lengths were 
measured with a fixed redshift distribution for each generated amplitude. 
Finally, the dispersion of calculated correlation lengths was assigned to be the 
uncertainty. 

\section{Results}

% figure : colour magnitude relation
\begin{figure}
\includegraphics[width=8cm]{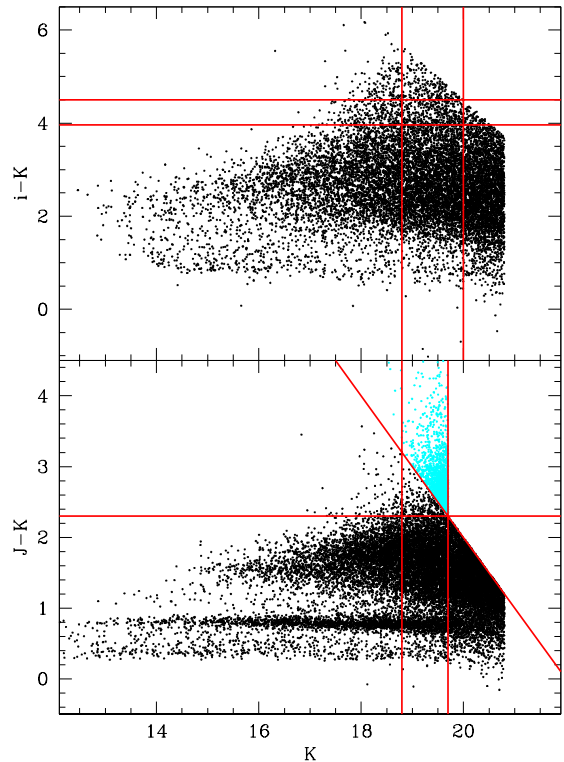}
\caption{The colour-magnitude diagrams for EROs (top) and DRGs (bottom).
The lines indicate the selection criteria for each population. For 
display purpose, only 10 per cent of all detected objects 
for $iK$ and $JK$ diagrams were displayed. The open circles are DRG candidates 
having $J>22$.}
\end{figure}

% figure : number density
\begin{figure}
\includegraphics[width=8cm]{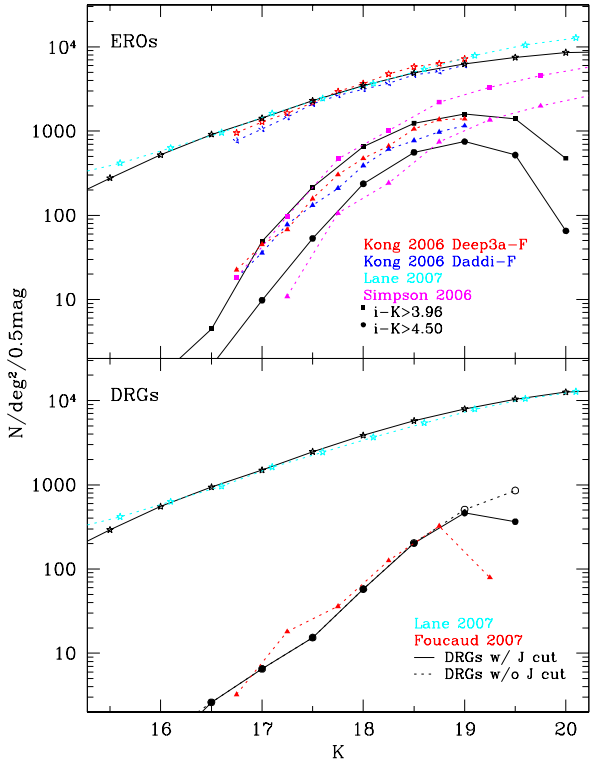}
\caption{Number counts of all galaxies (upper lines in each panel), EROs (top) 
and DRGs (bottom). The top panel 
shows the number counts of Deep3a-F (filled triangle) and Daddi-F (filled 
square) in Kong et al. (2006), $R-K>5.3$ (open square) and $R-K>6.0$ (open 
triangle) EROs in Simpson 
et al. (2006), all galaxies (asterisk) in Lane et al. (2007) and this work 
(star). 
The results of two selection criteria are presented by open circle 
($i-K>3.96$) and filled circle ($i-K>4.5$) symbols.
The bottom panel shows all galaxies (asterisk) in Lane et al. (2007), 
DRGs (triangle) in Foucaud et al. (2007) and DRGs with a $J$ magnitude cut 
(black filled circle) and without a
$J$ magnitude cut (black open circle), see text for details. Those of all galaxies in this 
work are results from $iK$ matched (top panel) and only $K$ (bottom panel) 
samples.}
\end{figure}

\subsection{Colour selection and Number Counts}

In this study, EROs and DRGs were selected using various colour criteria. 
Firstly, to remove the large majority of the Galactic stars from the optical-
IR catalogue, we used $g-J=33.33(J-K)-27$ for $J-K>0.9$ and $g-J>3$, and 
$g-J=4(J-K)-0.6$ for $J-K<0.9$ and $g-J<3$ introduced by 
Maddox et al. (2008). A $J-K<1$ criterion was used for the near-IR 
catalogue to remove all potential stars. 
Although many 
colour criteria for EROs exist, the redshift distribution of EROs in previous 
studies showed $z\sim$0.8 as a low redshift limit (Simpson et al. 2006; 
Conselice et al. 2008). The variation in $i-K$ with redshift predicted from the model galaxy SED 
in Kong et al. (2009) and the photometric redshift distribution we
find from the NMBS indicate that $i-K>4.5$ is appropriate to select 
$z>1$ galaxies. Therefore, $i-K>4.5$ was applied to select EROs in
keeping with comparable studies but we investigate the impact of
varying this cut in section 5. 
Similarly, we use $J-K>2.3$ to select DRGs. 
Due to the limitations
in the CTIO and DXS imaging our analysis is limited to
$i<24.6$ and $J<22.0$. So our absolute limit for selecting
EROs is $K<20.0$ and DRGs is $K<19.7$, well within our
completeness in $K$.
Each ERO  and DRG requires a joint detection in $i$ and $K$ or
$J$ and $K$ respectively, although we do investigate the 
number of possible EROs and DRGs where no detection
is found in the bluer band.
Figure 4 presents colour-magnitude diagrams for 
EROs (top) and DRGs (bottom) with selection criteria (lines). In the top 
panel, horizontal lines are $i-K=3.96$ (corresponding to $I-K=4.0$) and $i-K=4.5$, 
and the vertical lines are the $K=$18.8 and 20.0 magnitude 
limits. In addition, lines in the bottom panel indicate the $J-K=2.3$ and 
$K=$ 18.8 and 19.7 limits. The open circles are DRG candidates having $J>22$. 
Finally, 5,383 EROs and 3,414 DRGs with matched detections were selected.

Figure 5 shows the number counts of EROs (top) and DRGs (bottom).
The upper lines in each panel are number counts of all galaxies.
In the top panel of figure 5, results from the UKIDSS UDS 
(asterisk, Lane et al. 2007), Deep3a-F (filled triangle) and Daddi-F (filled 
square) from Kong et al. 
(2006) and EROs with $R-K>$ 5.3 and 6 (open square and triangle) from Simpson 
et al. (2006) were also plotted for comparison. 
The number counts of all galaxies in SA22 field are in agreement with previous 
results. However, all galaxies with matching $i$ and $K$ detections (top) show slightly lower density 
at faint magnitudes, since our $i$ depth is not sufficient to cover the full near-IR depth.
Similarly, our ERO counts are slightly below those from previous results 
because of our relative depth in $i$. The filled and open circles 
indicate EROs in SA22 selected by $i-K>4.5$ and $i-K>3.96$, 
respectively. The filled triangle and square are results for EROs selected by 
$R-K>5$ from Deep3a-F and Daddi-F of Kong et al. (2006) and,  by open square 
and triangle, $R-K>5.3$ and $R-K>6$ EROs from UKIDSS UDS in Simpson et al. 
(2006). Our $i-K>3.96$ ERO counts are comparable to those of $R-K>5$-5.3 EROs
and our $i-K>4.5$ ERO counts match those of $R-K>6$ EROs. 
See section 5.4 for a discussion of how the colour selection affects
the clustering.

The bottom panel of figure 5 shows the number counts of DRGs. The results for 
all galaxies with a joint detection of $J$ and $K$
(stars) are in agreement with Lane et al. (2007). 
In addition, the counts of DRGs are also same as those of the
UKIDSS UDS EDR in Foucaud et al. (2007). We also plot the number counts of 
DRGs irrespective of whether there is a matched detection in $J$ (open circles)
which only shows a significant difference for the faintest bin.

Table 1 lists the number counts of each population. We note that the number 
counts of all galaxies are from only the $K$ band of the UKIDSS DXS 
catalogue without a $J$ limit. However, those for EROs and DRGs are limited by 
the $i$ and $J$ bands, especially at faint magnitudes.

% Table : number density of ERO and DRG
\begin{table}
\caption{The number counts in 
log[N($deg^{-2}$0.5$mag^{-1}$)] of 
galaxies $i-K>$3.96 and 4.5 EROs and DRGs. The number counts of all galaxies are
measured using only the $K$ magnitude from the UKIDSS DXS catalogue without a $J$ limit, 
but those for the EROs and DRGs are limited by the $i$ and $J$ depths.}
\centering
\begin{scriptsize}
\begin{tabular}{ccccc}\\ \hline
$K$ bin & galaxies& $i-K>3.96$ EROs & $i-K>4.5$ EROs & DRGs \\
\hline
15.0  & 2.165 &   -   &   -   &   -   \\
15.5  & 2.465 &   -   &   -   &   -   \\
16.0  & 2.742 &   -   &   -   &   -   \\
16.5  & 2.973 & 0.652 & 0.213 & 0.416 \\
17.0  & 3.174 & 1.690 & 0.991 & 0.814 \\
17.5  & 3.393 & 2.333 & 1.725 & 1.185 \\
18.0  & 3.586 & 2.815 & 2.374 & 1.761 \\
18.5  & 3.759 & 3.092 & 2.747 & 2.306 \\
19.0  & 3.900 & 3.203 & 2.875 & 2.665 \\
19.5  & 4.017 & 3.146 & 2.716 & 2.561 \\
20.0  & 4.101 & 2.678 & 1.815 &   -   \\ \hline
\end{tabular}
\end{scriptsize}
\end{table}

\subsection{Clustering of EROs}

% figure : ERO fit
\begin{figure}
\includegraphics[width=7.5cm]{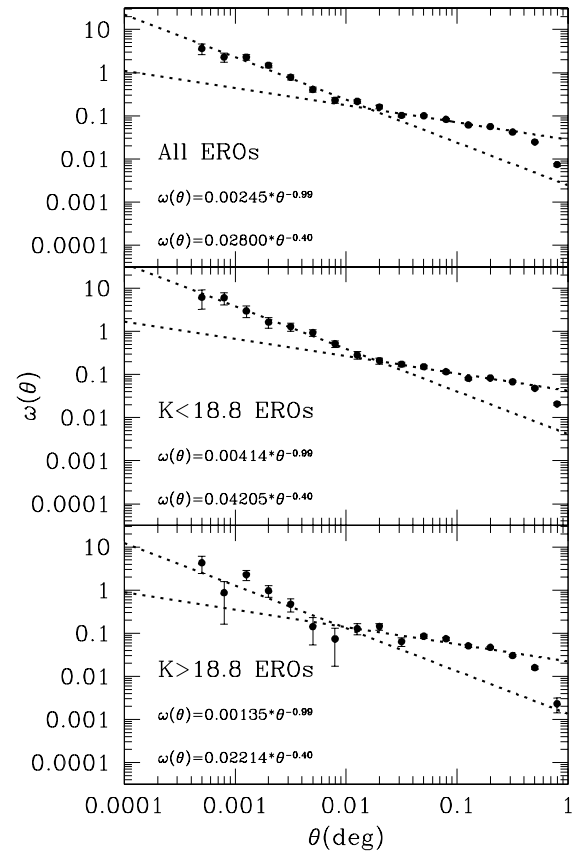}
\caption{The integral constraint corrected angular two-point correlation functions of all 
(top), $K<18.8$ (middle) and $K>18.8$ (bottom) EROs. Dotted lines and 
equations in each panel show power law ($A_{\omega}\theta^{-\delta}$) fitting 
results at small and large scale.}
\end{figure}

% Table : amplitude of ERO and DRG
\begin{table*}
\caption{The amplitudes $A_{\omega}$ of the correlation functions 
and the number of selected objects in the ERO and DRG samples.}
%\label{amplitude}
\centering
\begin{scriptsize}
\begin{tabular}{ccccccccc}\\ \hline
Criterion & && EROs & && &DRGs &\\
 && $A_{\omega}^{small}$($\times$10$^{-3})$  & $A_{\omega}^{large}$($\times$10$^{-3}$) & Num && $A_{\omega}^{small}$($\times$10$^{-3}$) &$A_{\omega}^{large}$($\times$10$^{-3}$) & Num\\
\hline
All       && 2.45$\pm$0.13 & 28.00$\pm$0.39 & 5,383 && 0.19$\pm$0.02 & 7.30$\pm$0.44 & 3,414\\
$K<18.8$  && 4.14$\pm$0.33 & 42.05$\pm$0.93 & 2,277 && 0.37$\pm$0.10 &23.63$\pm$1.57 & 979\\
$K>18.8$  && 1.35$\pm$0.20 & 22.14$\pm$0.67 & 3,106 && 0.16$\pm$0.03 & 6.23$\pm$0.67 & 2,435\\ \hline
\end{tabular}
\end{scriptsize}
\end{table*}

Figure 6 shows the  angular two-point correlation function corrected for the integral constraint of 
all (top), $K<18.8$ (middle) and $K>18.8$ (bottom) EROs. 
Many studies have found EROs are strongly clustered (Roche et al. 2002, 2003; 
Brown et al. 2005; Kong et al. 2006, 2009). A
power law fit to the angular correlation function 
is consistent with the previous data with no significant differences
in the measured slope. However, all of these studies has been restricted
to relatively narrow fields ($<20'$ on a side) and small samples ($<500$).
In this study, we 
represent the correlation function of EROs on much wider scales 
and with a larger sample than before.
We confirm
that EROs are strongly clustered at all of the angular scales sampled. 
There is an apparent 
inflection at $\theta\sim$ 0.6$'$-1.2$'$ implying a double power law is 
required to fit the correlation function of EROs. This has 
been observed for LRGs (Ross et al. 2008; Sawangwit et al. 2010) 
and DRGs at $2<z<3$ (Quadri et al. 2008) and is naturally explained by the 1- and 2-halo 
terms arising in the halo model of galaxy clustering.
We fit the correlation function 
of EROs by separating small ($\theta<$ 0.76$'$) and 
large (0.76$'<\theta<$ 19$'$) scales, and apply these ranges for all other 
ERO samples. The slopes of each power 
law were measured from $K<18.8$ and $i-K>4.5$ ERO samples by the fit 
described in section 3.4, and then those were applied to all and 
$K>18.8$ EROs, since our $i$-band magnitude limit prevents a complete
extraction of the faint EROs. The measured slopes are 0.99$\pm$0.09 for the 
small scales and 0.40$\pm$0.03 for the larger scales. The slopes are slightly 
smaller than those for LRGs at $z<1$ (1.16$\pm$0.07 and 0.67$\pm$0.07 of 2SLAQ and 
1.28$\pm$0.04 and 0.58$\pm$0.09 of AA$\Omega$ for small and large scale 
respectively) in Ross et al. (2008). However, our values are in agreement with DRGs 
at $2<z<3$ (1.2$\pm$0.3 and 0.47$\pm$0.14 for small and large scale 
respectively) in Quadri et al. (2008) within the uncertainty ranges. Also bright 
EROs show a larger amplitude than others, i.e., stronger clustering. The second 
and third columns in table 2 list the amplitudes measured. 

Perhaps the most striking result is the very shallow slope of
the clustering on the larger scales. The value of 0.4--0.5 is in
stark contrast to the canonical value of 0.8 that is so widely
found in lower redshift studies and assumed for more distant 
studies when the slope is poorly constrained. 
For our ERO and DRG samples, the DRGs in  Quadri et al. (2008) and the FIR selected galaxies in
Cooray et al. (2010), the strongly clustered objects are being
compared in projection over a range of order unity in redshift and
all show relatively shallow slopes on scales equivalent to
5--50~$h^{-1}$~Mpc.
This contrasts with the equivalent angular clustering of LRGs
by Sawangawit et al. (2010) where the depth in redshift is
at most 0.2 and the slope is steeper. 
A similar change in the slope has been 
noted in faint galaxy clustering by Neuschaefer \& Windhorst (1995) and
Postman et al. (1998) in which they parameterised the 
change in slope as $\delta(z)=1.75-1.8 (1+z)^{-0.2-0.35} -1$, 
where $z$ is the median redshift of the galaxies sampled.
This functional form is consistent with the shallower slope
we find for EROs compared to LRGs as long as the index
is less than -0.3, although this would predict a much
shallower slope at higher redshifts which would appear
to be inconsistent with results.

The origin of
this change in slope in the angular clustering is a combination
several factors. The primary one is the fact that
the angular diameter distance at redshifts above 0.8
is relatively constant, although this
is strongly dependent on the
cosmological parameters as shown by Kauffmann et al. (1999)
from N-body simulations. Indeed, the angular clustering
as a function of redshift may be a relatively simple
test of the Cosmological Constant and has been
proposed as a method to detect the Baryon Accoustic Oscillation
scale by S\'{a}nchez et al. (2010).
Another effect that leads to the flattening of
the slope may also be the redshift range
sampled as the clustering is diluted as galaxies
of differing distances are compared. We note that
S\'{a}nchez et al. (2010) predict a significantly
shallower slope for the angular clustering
as the redshift range increases on the
scales we are considering here.
Future studies
will be able to test this directly with improved
photometric redshift accuracy.

Our results need to be compared to previous results with 
careful attention to the differences in our selection and measurement. 
Thus we applied the same method with the previous studies which 
measured the integral constraint and the amplitude of correlation function by a 
single power law with the fixed slope as $\delta=0.8$.
First we consider the amplitude of the clustering which may
appear to differ only because of the fact we are fitting a double
power law. Fitting a single power law to the angular clustering of 
$K<18.8$ and $i-K>4.5$ EROs 
we measure an amplitude of (12.72$\pm$0.5)$\times10^{-3}$ which is 
consistent with (14.60$\pm$1.64)$\times10^{-3}$ of Daddi-F EROs at the same 
magnitude limit, but slightly larger than (9.29$\pm$1.60$)\times10^{-3}$ of 
Deep3a-F EROs in Kong et al. (2006). In addition, our value is larger than 
(6.6$\pm$1.1)$\times10^{-3}$ of Kong et al. (2009). However,
this difference may be the result of the different selection criteria 
and angular ranges used.
To illustrate this, if we fit over the angular range sampled by
Kong et al. (2009) of 0.19$'$ and 3$'$
to measure the amplitude of $i-K>3.96$ and $K<18.8$ EROs with a single 
power law with $\delta=$ 0.8 we recover a value of 
(6.65$\pm$0.3)$\times10^{-3}$ that does match their published value. 
Therefore, our results are
entirely consistent with Kong et al. (2009) given the effects of 
Cosmic Variance (see section 5.6) even though the slopes and
amplitudes we quote appear to differ on first inspection.
 
Secondly, if a single 
power law is applied to fit the correlation function, the reduced $\chi^{2}$ 
value is 2.8. However, the value drops to 0.3 for small scales and 1.5 for large 
scales, when the double power law with the measured slopes is applied. Thus a 
double power law well describes the correlation function of our EROs
but past observations have not uncovered it due to their limited angular
sampling and larger errors.

The clustering properties as a function of limiting magnitude and colour are discussed in 
section 5.

% Table : amplitude of magnitude limited EROs with fixed slope
\begin{table*}
\caption{The amplitudes $A_{\omega}$ of the correlation functions, 
correlation lengths with fixed slopes and number of EROs 
for each magnitude limit.}
%\label{amplitude}
\centering
\begin{scriptsize}
\begin{tabular}{ccccccc}\\ \hline
$K$ limit & $A_{\omega}^{small}$ ($\times$10$^{-3}$)  & $A_{\omega}^{large}$ ($\times$10$^{-3}$)& r$_{0}$$^{small}$ ($h^{-1}$ Mpc) & r$_{0}$$^{large}$ ($h^{-1}$ Mpc) & $\chi^{2}_{small, large}$ & Num.\\
\hline
$K<18.3$ & 7.52$\pm$1.1 & 42.06$\pm$2.5 & 14.09$\pm$1.9 & 21.51$\pm$1.3 &0.9, 2.1&852\\
$K<18.5$ & 5.90$\pm$0.6 & 38.18$\pm$1.6 & 12.85$\pm$0.7 & 20.92$\pm$0.6 &0.3, 2.8&1,323\\
$K<18.8$ & 4.14$\pm$0.3 & 42.05$\pm$0.9 & 11.29$\pm$0.5 & 23.97$\pm$0.4 &0.3, 1.5&2,277\\
$K<19.0$ & 3.79$\pm$0.3 & 41.86$\pm$0.7 & 11.12$\pm$0.4 & 24.89$\pm$0.3 &3.2, 2.0&3,014\\
$K<19.5$ & 2.80$\pm$0.2 & 31.69$\pm$0.5 &  9.96$\pm$0.3 & 21.60$\pm$0.2 &2.0, 3.1&4,713\\
$K<20.0$ & 2.45$\pm$0.1 & 28.00$\pm$0.4 &  9.48$\pm$0.3 & 20.29$\pm$0.2 &1.9, 3.6&5,383\\
\hline
\end{tabular}
\end{scriptsize}
\end{table*}
% Table : amplitude of magnitude limited EROs with variable slope
\begin{table*}
\caption{The amplitudes $A_{\omega}$ of the correlation functions, 
correlation lengths with variable slopes and estimated variable 
slopes of EROs for each magnitude limit.}
%\label{amplitude}
\centering
\begin{scriptsize}
\begin{tabular}{cccccccc}\\ \hline
$K$ limit & $A_{\omega}^{small}$ ($\times$10$^{-3}$)  & $A_{\omega}^{large}$ ($\times$10$^{-3}$)& r$_{0}$$^{small}$ ($h^{-1}$ Mpc) & r$_{0}$$^{large}$ ($h^{-1}$ Mpc) & slope$^{small}$ & slope$^{large}$ & $\chi^{2}_{small, large}$\\
\hline
$K<18.3$ & 3.20$\pm$0.4 & 33.26$\pm$1.9 & 10.51$\pm$1.4 & 21.97$\pm$1.5 &1.15$\pm$0.15 &0.53$\pm$0.07&0.7, 1.8\\
$K<18.5$ & 4.05$\pm$0.4 & 34.92$\pm$1.5 & 11.28$\pm$0.6 & 21.27$\pm$0.6 &1.06$\pm$0.12 &0.45$\pm$0.05&0.2, 2.7\\
$K<18.8$ & 4.14$\pm$0.3 & 42.05$\pm$0.9 & 11.29$\pm$0.5 & 23.97$\pm$0.4 &0.99$\pm$0.09 &0.40$\pm$0.03&0.3, 1.5\\
$K<19.0$ & 3.40$\pm$0.2 & 44.14$\pm$0.8 & 10.71$\pm$0.4 & 24.59$\pm$0.3 &1.01$\pm$0.07 &0.37$\pm$0.02&0.4, 1.8\\
$K<19.5$ & 3.31$\pm$0.2 & 28.98$\pm$0.4 & 10.54$\pm$0.3 & 21.94$\pm$0.2 &0.96$\pm$0.06 &0.45$\pm$0.02&2.0, 2.1\\
$K<20.0$ & 2.45$\pm$0.1 & 25.14$\pm$0.4 &  9.60$\pm$0.3 & 20.70$\pm$0.2 &0.99$\pm$0.05 &0.46$\pm$0.02&1.9, 2.3\\
\hline
\end{tabular}
\end{scriptsize}
\end{table*}

\subsection{Clustering of DRGs}

% figure : DRG fit
\begin{figure}
\includegraphics[width=7.5cm]{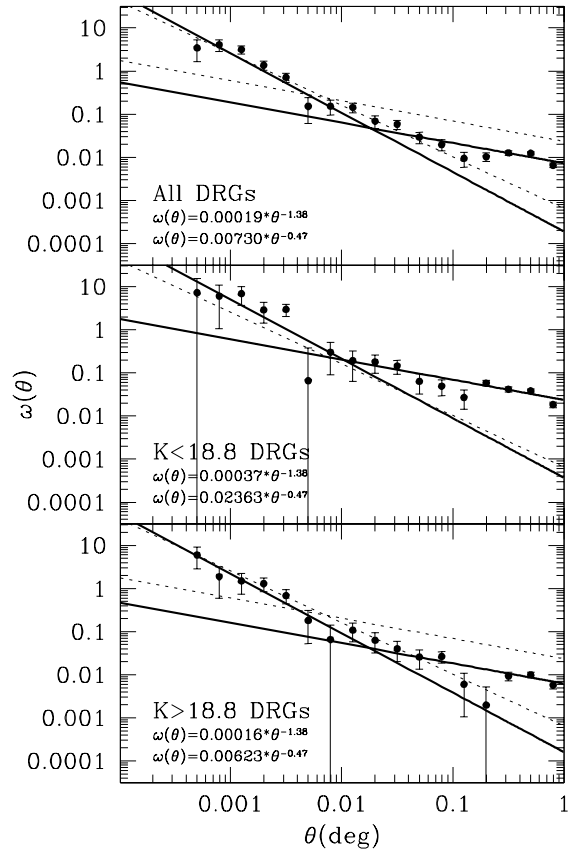}
\caption{The angular two-point correlation functions corrected for the integral constraint of all 
(top), $K<18.8$ (middle) and $K>18.8$ (bottom) DRGs. Solid lines and 
equations in each panel show power law fitting results. Dotted lines are the 
best fit results for $2<z<3$ DRGs from Quadri et al. (2008)}
\end{figure}

Applying a similar analysis to the angular clustering of DRGs, we again find that
a double power law fit is required (Figure 7).
While most early attempts to measure the angular correlation
of DRGs were consistent with a single power law (Grazian et al. 2006; 
Foucaud et al. 2007), Quadri et al. (2008) demonstrated 
that a double power law with an inflection at 
$\theta\sim$0.17$'$ was appropriate to fit the angular correlation function 
of $2<z<3$ DRGs in the UKIDSS UDS field. 
However, the angular ranges for the
small and large scales used in their fitting were 
split at 0.67$'$. To ensure that our results can be
compared, we have used the power law 
slope for the large scale clustering found by Quadri et al. (2008), $\delta=$0.47, 
and measured that for the small scales from a free
fit to $\delta$ after the integral constraint correction.

In order to measure an amplitude and slope for each angular range, small and 
large scales were split at 0.48$'$ since our correlation functions showed an 
upturn at $\theta<$ 0.48$'$. Then the power law slope for small scales was 
measured for $K<18.8$ DRGs which is not affected by the $J$ magnitude limit. 
The measured slope was $\delta=$ 1.38$\pm$0.27, which is consistent with the
value of 1.2$\pm$0.3 derived by Quadri et al. (2008) 
considering the additional photometric redshift constraint they applied.

To directly compare our angular correlation function of DRGs with 
that in Foucaud et al. (2007), the function for $K<18.8$ DRGs
was fitted with a single power law with a fixed $\delta$ of 1.0 between 0.5$'$ and 12$'$
to match their magnitude limit and fit constraints.
The amplitude of $K<18.8$ DRGs, 
(3.07$\pm$0.6)$\times$10$^{-3}$, is consistent with 
3.1$^{+2.1}_{-1.3}$$\times$10$^{-3}$ in Foucaud et al. (2007). 

% figure : ERO limited magnitudes & fit
\begin{figure}
\includegraphics[width=7.5cm]{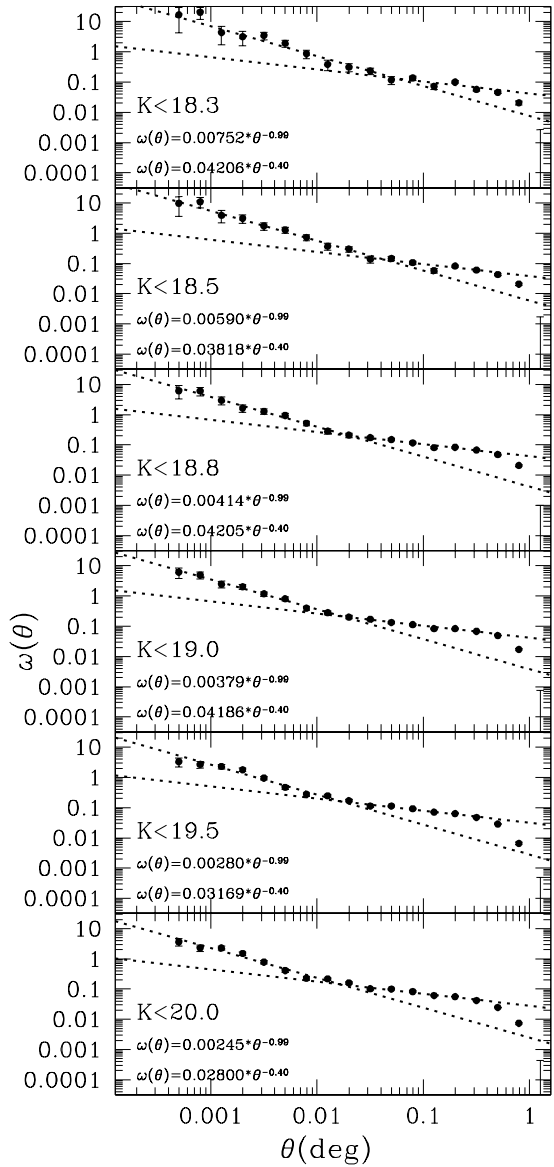}
\caption{The angular two-point correlation functions corrected for the integral constraint and 
fitted power laws for various magnitude limited samples of EROs.}
\end{figure}

Figure 7 presents the  angular correlation functions corrected for the integral constraint for all 
(top), $K<18.8$ (middle) and $18.8<K<19.7$ (bottom) DRGs with fitted power laws. 
The solid and dotted lines are the fitted power law and that for $2<z<3$ DRGs 
in Quadri et al. (2008), respectively. It is apparent that DRGs are strongly 
clustered, and their correlation functions are well described by a double 
power law. There are clear differences in the amplitude of 
clustering on both small and large scales between our brighter and
fainter DRGs and in the angular scale for the inflection compared to the 
Quadri et al. (2008) sample. The measured amplitudes are 
listed in table 2 but should be used with care given the complex
interplay between the depth, redshift sampled and angular coverage of 
DRG samples.

\section{Discussion}

% Table : amplitude of colour limited EROs with fixed slope
\begin{table*}
\caption{The amplitudes $A_{\omega}$ of the correlation functions, 
the correlation lengths with fixed slopes and the number of EROs 
for each colour limit.}
%\label{amplitude}
\centering
\begin{scriptsize}
\begin{tabular}{ccccccc}\\ \hline
$i-K$ limit & $A_{\omega}^{small}$ ($\times$10$^{-3}$)  & $A_{\omega}^{large}$ ($\times$10$^{-3}$)& r$_{0}$$^{small}$ ($h^{-1}$ Mpc) & r$_{0}$$^{large}$ ($h^{-1}$ Mpc) & $\chi^{2}_{small, large}$ & Num.\\
\hline
$i-K>3.96$ & 2.32$\pm$0.1 & 21.92$\pm$0.4 &  8.63$\pm$0.2 &15.75$\pm$0.2&3.2, 5.6& 5,654\\
$i-K>4.3$  & 3.24$\pm$0.2 & 33.87$\pm$0.6 & 10.24$\pm$0.4 &21.39$\pm$0.3&0.6, 3.7& 3,313\\
$i-K>4.5$  & 4.14$\pm$0.3 & 42.05$\pm$0.9 & 11.29$\pm$0.5 &23.97$\pm$0.4&0.3, 1.5& 2,277\\
$i-K>4.8$  & 4.52$\pm$0.6 & 55.19$\pm$1.4 & 11.66$\pm$0.9 &28.48$\pm$0.5&0.8, 5.0& 1,259\\
\hline
\end{tabular}
\end{scriptsize}
\end{table*}
% Table : amplitude of colour limited EROs with variable slope
\begin{table*}
\caption{The amplitudes $A_{\omega}$ of the correlation functions, 
the correlation lengths with variable slopes and the estimated 
slopes of EROs for each colour limit.}
%\label{amplitude}
\centering
\begin{scriptsize}
\begin{tabular}{cccccccc}\\ \hline
$i-K$ limit & $A_{\omega}^{small}$ ($\times$10$^{-3}$)  & $A_{\omega}^{large}$ ($\times$10$^{-3}$)& r$_{0}$$^{small}$ ($h^{-1}$ Mpc) & r$_{0}$$^{large}$ ($h^{-1}$ Mpc) & slope$^{small}$ & slope$^{large}$&$\chi^{2}_{small, large}$\\
\hline
$i-K>3.96$ & 2.19$\pm$0.1 & 17.95$\pm$0.3 &  8.46$\pm$0.2 &16.41$\pm$0.2&1.00$\pm$0.05 &0.51$\pm$0.02&3.2, 2.2\\
$i-K>4.3$  & 3.43$\pm$0.2 & 29.32$\pm$0.6 & 10.45$\pm$0.4 &21.81$\pm$0.3&0.98$\pm$0.07 &0.48$\pm$0.02&0.6, 2.3\\
$i-K>4.5$  & 4.14$\pm$0.3 & 42.05$\pm$0.9 & 11.29$\pm$0.5 &23.97$\pm$0.4&0.99$\pm$0.09 &0.40$\pm$0.03&0.3, 1.5\\
$i-K>4.8$  & 2.79$\pm$0.4 & 57.52$\pm$1.4 &  9.92$\pm$0.7 &27.97$\pm$0.5&1.08$\pm$0.17 &0.37$\pm$0.03&0.8, 4.8\\
\hline
\end{tabular}
\end{scriptsize}
\end{table*}

% figure : amplitudes of EROs with various limit mag
\begin{figure}
\includegraphics[width=7.5cm]{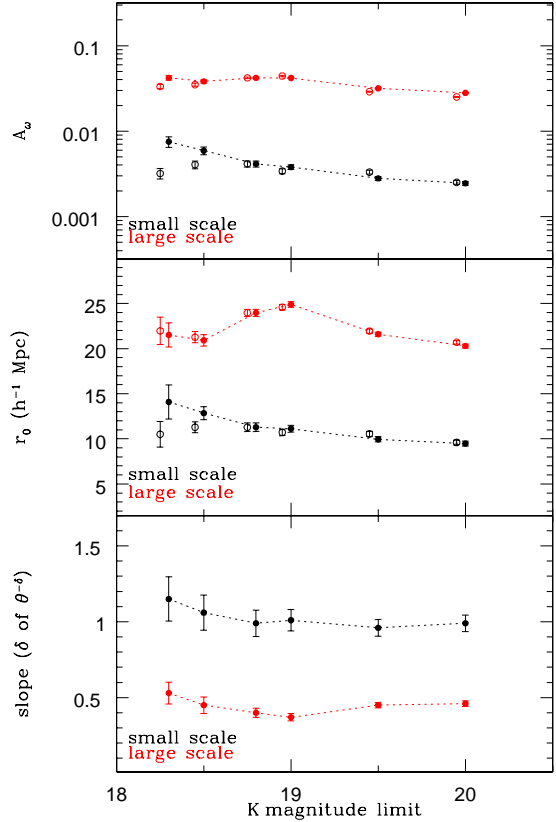}
\caption{The amplitudes (top) and real space correlation lengths (middle) of 
double power law fits with fixed slopes ($\delta=$0.99,0.40) for magnitude 
limited EROs and the measured slope (bottom) as a function of limiting magnitude.
The open symbols in the top and middle panels are results when the slope is allowed to vary.
Those are slightly shifted for display purposes.}
\end{figure}

\subsection{Magnitude limited EROs}

It is well known that the clustering of EROs depends on the limiting magnitude 
of any selection criterion (Daddi et al. 2000; Roche et al. 2002, 2003; 
Brown et al. 2005; Georgakakis et al. 2005; Kong et al. 2006, 2009). 
However, in previous cases, a single power law was invoked
to describe the angular correlation function. In this section, we discuss 
the clustering properties at small and large scales with various magnitude 
limits to expand on these previous studies.

To select EROs at each magnitude limit, the colour was fixed at $i-K>4.5$. 
We applied the same slopes of power law and fitting ranges used in 
section 4.2 for small and large scales, and fitted 
$\omega(\theta)=A_{\omega}\theta^{-\delta}$ to 
the correlation function. Figure 8 shows the angular 
two-point correlation functions corrected for the integral constraint
and fitted results for various subsets of EROs using 
different limiting magnitudes with fixed colour. It is apparent 
that a double power law is required to fit the correlation function of EROs 
for all magnitude limits.

The top panel in figure 9 shows the amplitude of each power law with fixed slopes 
as a function of limiting magnitude (filled symbols). Although faint EROs may 
not be complete because of the relatively shallow $i$ depth, there is a trend 
in the amplitude at the brightest limiting magnitudes. The amplitude of the 
small scale varies significantly. However, the amplitude of the large scale 
shows an almost constant value at all magnitude limits.

The variation in  amplitude at small scales is also apparent in the real space 
correlation length in the middle panel of figure 9 (filled symbols). As 
mentioned in section 
3.4, the amplitudes measured with fixed slopes from $i-K>4.5$ and $K<18.8$ 
EROs were used to calculate the correlation length. The correlation length for 
the small scales shows a range between 9 and 14 $h^{-1}$ Mpc with the strongest 
clustering for the brighter galaxies.
On the other hand, the clustering on large scales shows a 
similar length of  21--24 $h^{-1}$ Mpc that varies marginally over 
the range in magnitude sampled.

To investigate the variation in slope, we measured slopes by fitting a power law 
with a free slope. The results are presented in the bottom panel of figure 9. 
The slope of the brighter sub-samples have higher values than the fainter ones, 
i.e., brighter EROs show steeper correlation functions especially on small 
scales. The amplitudes and correlation lengths from freely fitted slopes are 
presented in the top and middle panel of figure 9 with open symbols. For
display purposes the points are slightly shifted in magnitude. All estimated values with fixed slopes 
are listed in table 3 and those with variable slopes are in table 4.

To compare to previous results we again need to
fit a single power law to a smaller range in angle.
We find the correlation length of $K<20$ and $i-K>4.5$ EROs 
with a fixed $\delta=$0.8 is 16.99$\pm$0.2 $h^{-1}$ Mpc, 
which is consistent with 12--17 $h^{-1}$ Mpc in Georgakakis et al. (2005).
Our value may be higher than Georgakakis et al. due to our redder
colour limit that preferentially selects more massive galaxies.
Furthermore, the correlation length of our $K<18.8$ 
and $i-K>3.96$ EROs fitted by a single power law (see section 4.2) is 
12.52$\pm$0.33 $h^{-1}$ Mpc which is higher than 9.6$\pm$0.1 or 9.2$\pm$0.2 
$h^{-1}$ Mpc in Kong et al. (2009), although the amplitudes are all consistent. 
This is most probably caused by the different redshift distribution of the Kong $R-K$ sample.
Applying the different criteria to the NMBS sample, our selection has a slightly larger fraction 
of galaxies at $1.5<z<2.0$ than that in Kong et al. (2009).

% figure : ERO limited colours & fit
\begin{figure}
\includegraphics[width=7.5cm]{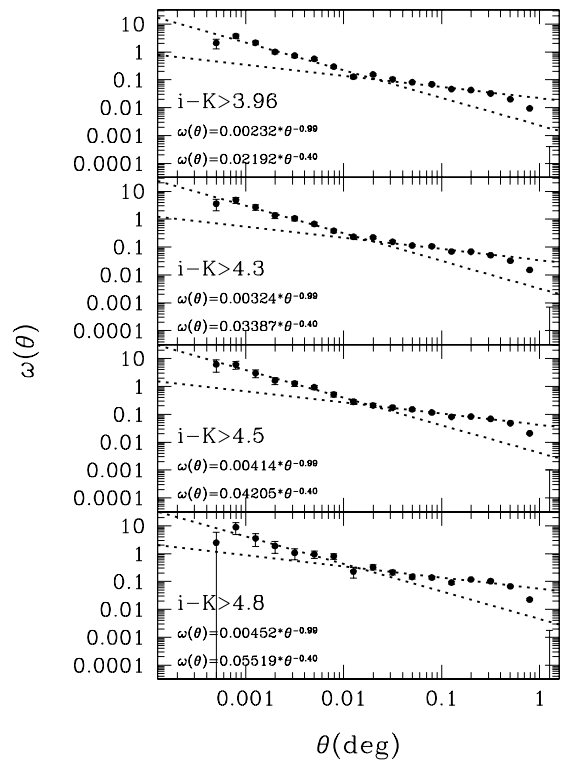}
\caption{The angular two-point correlation functions corrected for the integral constraint and 
fitted results for various colour limits.}
\end{figure}

\subsection{Colour limited EROs}

% figure : amplitudes of EROs with various limit colours
\begin{figure}
\includegraphics[width=7.5cm]{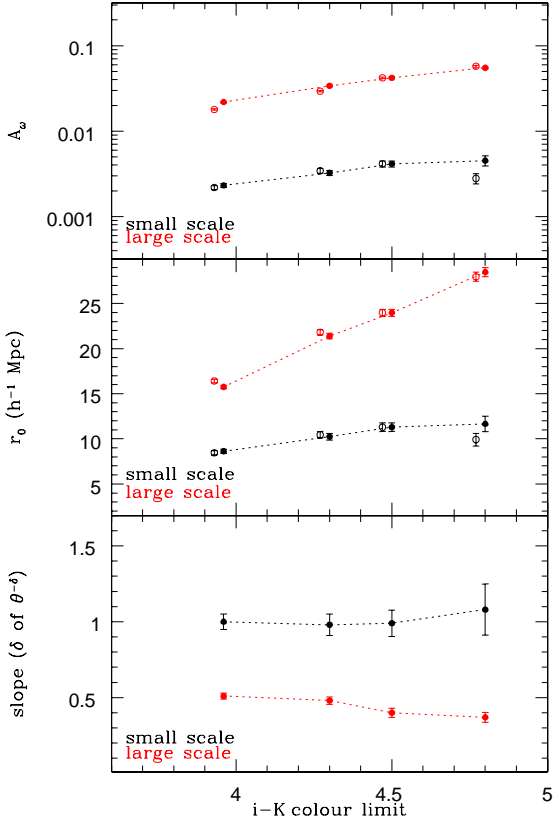}
\caption{The amplitudes (top) and real space correlation lengths (middle) of 
double power law fits with fixed slopes ($\delta=$0.99,0.40) for colour 
limited EROs and measured slope (bottom) as a function of colour limits.
The open symbols in the top and middle panels are measured from fits with a varying slope.
Those are slightly shifted for display purposes.}
\end{figure}

Daddi et al. (2000) studied the clustering amplitude as a function of 
colour threshold. They pointed out that red galaxies have a higher amplitude, 
but the amplitudes of $R-K>5.0$ and $R-K>5.3$ EROs were consistent. In 
addition, Brown et al. (2005) also noted no significant difference in 
amplitude of $R-K>$ 5.0 and 5.5 EROs. However, the small area or shallower depth 
of previous studies may have prevented a sufficiently accurate measurement of 
the ERO clustering to recover these differences. 
In this section we discuss the 
clustering properties of EROs as a function of colour limit using wider coverage than 
Daddi et al. (2000) and deeper imaging than Brown et al. (2005).

We used only $K<18.8$ ERO samples with various colour limits since the $i$ depth 
is too shallow to cover the full near-IR depth for the reddest sub-samples. 
Figure 10 shows 
angular correlation functions corrected for the integral constraint for various colour limits and the  
fitted power laws. The same ranges and fixed slopes from section 4.2 were applied 
to fit the functions. As verified in previous sections, the double power law 
well describes the functions of all colour limited EROs. However, 
the correlation functions show different trends from 
those of magnitude limited EROs. The relation between 
amplitude and colour limit is presented in the top panel of figure 11 (filled 
symbols). It is clear that redder EROs have higher amplitudes, i.e., stronger 
clustering. Moreover, there is a similar trend in the amplitude as a function 
of colour limit for both small and large scales. This is also evident in the 
trend of the real space correlation length in the middle panel of figure 11 
(filled symbols), but the
correlation lengths for large scales vary most dramatically. 
This increased clustering with colour limit is entirely as expected
given the correlation between colour and lower redshift limit of the
selection. The redder colour cuts select more distant, more luminous
galaxies that are therefore more clustered.
In the 
bottom panel of figure 11, the slopes measured independently show similar 
values for each scale at all colour limits. 
The lack of any variation with colour indicates that the 
form of the clustering does not change dramatically with redshift.
The values of the  freely fitted slope are also marked as open symbols in the
top and middle panels in figure 11. The measured values using a fixed slope are listed 
in table 5, and those with a variable slope are in table 6. 

\subsection{Populations of EROs}

% figure : two-colour diagram for OG and DG
\begin{figure}
\includegraphics[width=7.5cm]{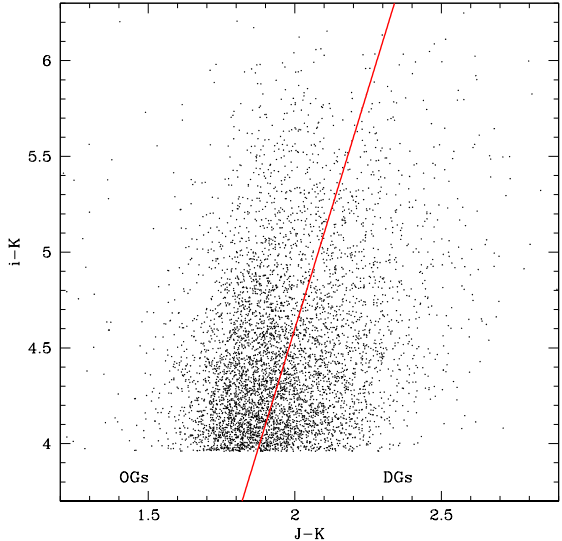}
\caption{The $i-K$ vs. $J-K$ colour-colour diagram for $K<18.8$ EROs. The line 
indicates the criterion defined by Fang et al. (2009) to classify OGs and DGs.}
\end{figure}

% figure : OG and DG
\begin{figure*}
\includegraphics[width=12cm]{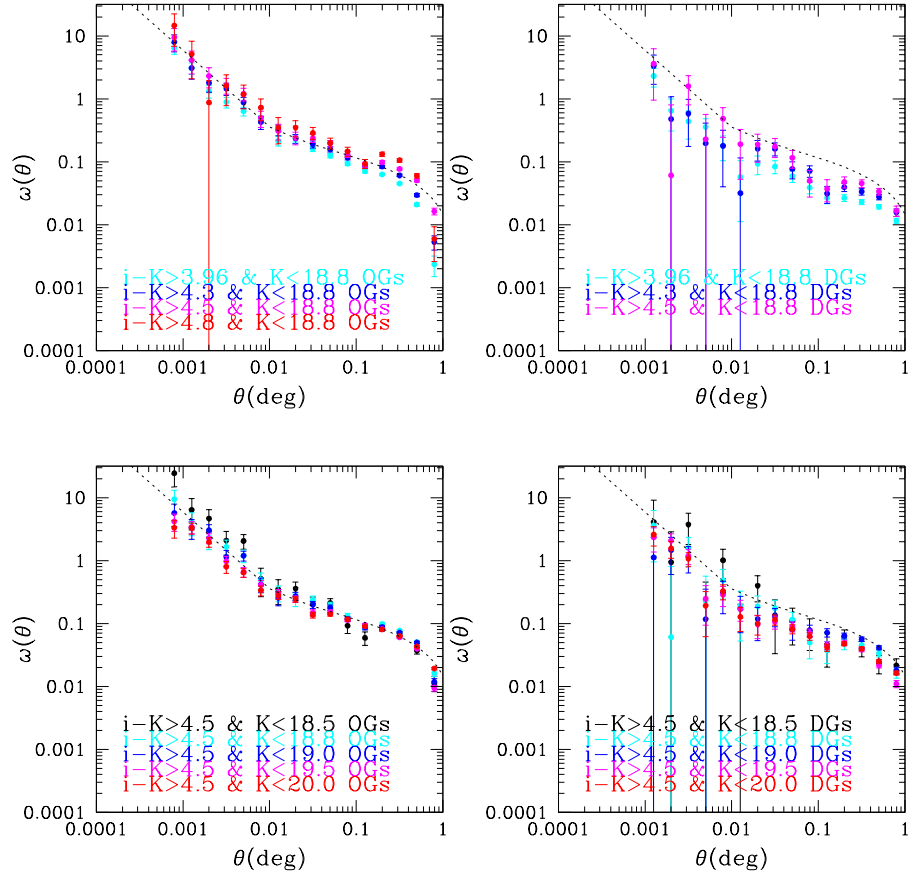}
\caption{The angular correlation functions corrected for the integral constraint of OGs (left) and 
DGs (right). Also shown are the correlation functions for magnitude limited 
(bottom) and colour limited (top) samples. The dotted line indicates the best fit for the 
correlation function of $K<18.8$ and $i-K>4.5$ EROs.}
\end{figure*}

The simple colour selection of EROs, while effective and easy
to implement, does not necessarily return a uniform population
of galaxies. 
It is known that EROs can be classified into old passively evolved galaxies 
(OGs) and dusty star-forming galaxies (DGs). Pozzetti \& Mannucci 
(2000) suggested a colour criterion in the ($I-K$) versus ($J-K$) plane 
defined by differences in the SEDs for old stellar populations and 
dusty galaxies. However, in this study, the $i$ filter was used instead of $I$. 
Therefore we adopted the criterion, ($J-K$)$=$0.20($i-K$)$+$1.08, defined 
in the ($i-K$) versus ($J-K$) plane by Fang, Kong \& Wang (2009). Figure 12 
shows the two-colour diagram of $K<18.8$ EROs for classifying OGs and DGs. 
The line indicates the criterion defined by Fang et al. (2009). The 
fraction of OGs with $K<18.8$ and $i-K>4.5$ is $\sim$63 per cent. This value 
is consistent with the fractions found for $K<19.7$ EROs selected by
$I-K$ and $R-K$ in Conselice et al. (2008). 
The fractions of DGs 
selected by various magnitude cuts with $i-K>4.5$ are 36 per cent at $K<18.5$ 
and 41 per cent at $K<20$, although this may be affected by the shallow 
$i$-band depth. The fraction of colour limited $K<18.8$ DGs are 43 per cent 
for $i-K>3.96$ EROs and 36 per cent for $i-K>4.8$ EROs. 

Figure 13 shows the angular correlation functions corrected for the integral constraint of OGs (left) 
and DGs (right). In addition, correlation functions were measured with various colour 
(top) and magnitude (bottom) limits. The dotted line of each panel indicates 
the best fit for the correlation function for $K<18.8$ and $i-K>4.5$ EROs. 
Since the small number of objects for $i-K>4.8$ DGs can cause a poor 
statistical and uncertain integral constraint, $i-K>4.8$ DGs were excluded 
for analysis.
The most apparent features are that the correlation functions for OGs show a clear break at 
the same position as that for all EROs in section 4.2, and OGs are more 
clustered than the full ERO sample as a function of both magnitude
and colour limit, especially on large scales.
Conversely, the correlation functions of DGs show much more scatter 
between sub-samples in magnitude and colour
and are much less clustered than OGs and the full ERO sample. This
can be attributed to their wider redshift range and lower intrinsic
mass.

These trends are also confirmed in the real space correlation lengths in 
figure 14. Figure 14 shows the correlation length of OGs (filled symbols) and 
DGs (open symbols) on small (circle) and large (triangle) scales plotted for 
various magnitude (bottom) and colour (top) limits. 
As the angular correlation function of some DG samples was not measured at 
$\sim$0.05$'$, to estimate the correlation 
length, the range between 0.076$'$ and 0.76$'$ for small scales (i.e.
narrower than that used earlier), was fitted with 
the fixed slopes used in section 4.2.
This may lead to slightly different correlation lengths at small 
scales, but it should not affect any overall 
trends.
The most important feature of the correlation lengths is the trend within each 
population. The magnitude limited OGs show no significant change in 
clustering on large scales and only a weak decline in small scale clustering
strength, as the uniformity of the correlation
functions in figure 13 implies. However redder OGs have larger correlation 
lengths than bluer ones, since redder OGs are more distant and more massive 
galaxies. Indeed, the similarity in the
clustering in strength and functional form to that of low
redshift LRGs in Sawangwit et al. (2010) implies that 
there is a continuity in the selection of massive, passive galaxies
that can be made from optical and near-IR surveys.
Similarly, the DGs show more significant
variation in clustering with colour and magnitude.

The consistency in clustering within the OG and DG samples
contrasts with the much more significant changes in clustering
seen in figures 9 and 11. These trends can be attributed
to the changes in the relative proportion of OGs and DGs 
as a function of colour and magnitude.
For instance, the decreased clustering on large scales for the
brightest and bluest EROs coincides with the largest 
fraction of DGs (up to 43 per cent) resulting in lower
clustering strength. These results highlight the need to treat
the selection of EROs with care as the diversity of
galaxies selected can lead to misleading clustering trends.

\subsection{EROs selected by $r-K$ colour}

Although various colour selection criteria have been used in
previous studies to select EROs, 
the differences between criteria have not been well characterised.
In this section, we briefly compare the clustering of  EROs selected by 
$r-K$ and $i-K$ colours to attempt to clarify how the use of a 
different optical filter affects their statistical and clustering properties.

Figure 15 shows the observed angular correlation functions of 5,564 $K<18.8$ EROs 
which are selected by $i-K>3.96$ (red), 4,326 EROs by $r-K>5$ and $r<24$ 
(the peak of our $r$-band number counts) (blue), 7,185 EROs by $r-K>5$ without 
$r$ magnitude cut (green) and 4,799 EROs by $r-K>5.3$ without $r$ magnitude 
cut (orange). It is clear that $r-K>5.3$ EROs (orange) are more clustered 
than $r-K>5$ EROs (green), matching the results found with
$i-K$ selection. 
However, the most apparent feature is that each sample shows 
different clustering properties, particularly on larger scales. The EROs 
selected by $i-K>3.96$ show the highest amplitude of all the samples and
the slope of the clustering on larger scales varies significantly with colour.
These different clustering properties are most easily explained by the 
changes in the redshift distribution between
samples and the different proportion of OGs to DGs.
Conselice et al. (2008) mentioned that the $I-K>4$ criterion is 
more useful to select EROs at higher redshift than $R-K>5.3$. In fact, 
$r$-band magnitudes of half of our $i-K>$3.96 EROs are fainter than our $r=$24 
magnitude limit so any $r-K$ sample would be incomplete.
Conversely, 27 per cent of the EROs selected by $r-K>5$ without an $r$ cut 
from our sample have $i-K<3.96$ and would hence not have been 
considered in any of our $i-K$ samples. 
These objects $r-K>5$ EROs that are blue in $i-K$ will be 
at lower redshifts, lower mass and therefore less clustered.

% figure : r0 of OGs and DGs
\begin{figure}
\includegraphics[width=7.5cm]{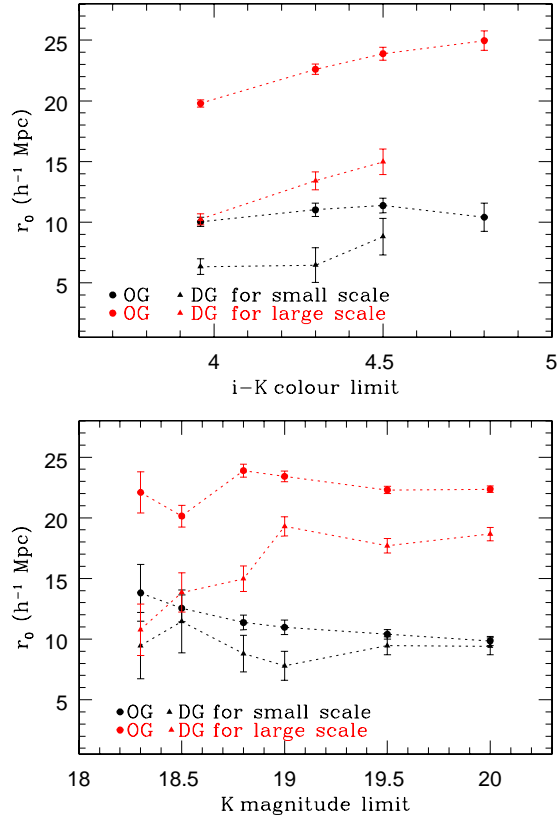}
\caption{The correlation length of OGs (filled symbols) and DGs (open symbols) 
for various magnitude (bottom) and colour (top) limits.}
\end{figure}

% figure : correlation function of r-K EROs
\begin{figure}
\includegraphics[width=7.5cm]{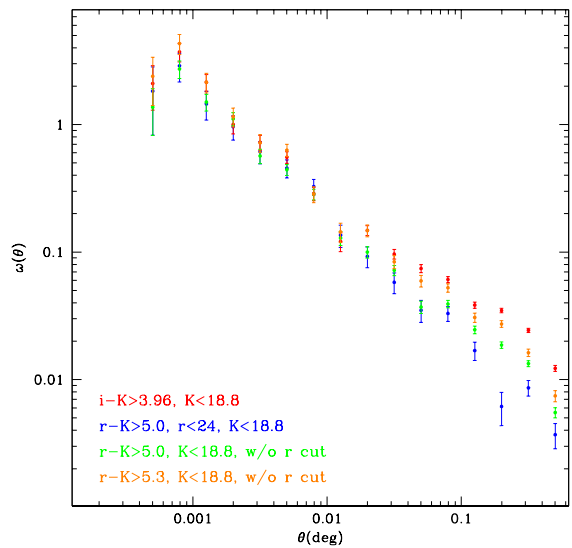}
\caption{The observed angular correlation functions of $K<18.8$ EROs with
$i-K>3.96$ (red), $r-K>5$ and $r<24$ (blue), $r-K>5$ without an $r$ cut 
(green) and $r-K>5.3$ without an $r$ cut (orange). It is noted that an integral
constraint correction has not been applied to these functions. }
\end{figure}

\subsection{Clustering of EROs and DRGs}

% Table : correlation lengths of ERO and DRG
\begin{table}
\caption{The correlation length r$_{0}$ ($h^{-1}$ Mpc) of selected objects for 
EROs and DRGs.}
%\label{amplitude}
\centering
\begin{scriptsize}
\begin{tabular}{ccccc}\\ \hline
Criterion & r$_{0,ERO}^{small}$&r$_{0,ERO}^{large}$&r$_{0,DRG}^{small}$&r$_{0,DRG}^{large}$\\
\hline
All       &9.48$\pm$0.3&20.29$\pm$0.2&4.66$\pm$0.2&10.32$\pm$0.4\\
$K<18.8$  &11.29$\pm$0.5&23.97$\pm$0.4&5.14$\pm$0.6&17.19$\pm$0.8\\
$K>18.8$  &7.12$\pm$0.6&17.45$\pm$0.4&4.42$\pm$0.4&9.52$\pm$0.7\\ \hline
\end{tabular}
\end{scriptsize}
\end{table}

 The goal of the colour criteria for EROs and DRGs are to select 
red galaxies that are likely to be at high redshift ($z>1$ or $z>2$).
However, Lane et al. (2007) and Quadri et al. (2007) 
find that the colour cut for DRGs can include a significant fraction of
relatively low redshift objects ($0.8<z<1.4$) that are dust obscured. 
In this section we briefly discuss the comparison of two different 
populations with clustering properties.

Table 7 lists correlation lengths of $i-K>4.5$ EROs and $J-K>2.3$ DRGs. It is apparent 
that brighter samples show stronger clustering. Although the direct comparison 
of correlation length is difficult because of different slopes and redshift 
distributions, we can confirm that EROs are more clustered than DRGs from the 
comparison of figure 6 and 7, and correlation lengths. This was also found by 
Foucaud et al. (2007). 

However, it has been shown that the $J-K>2.3$ criterion selects $1<z<2$ 
objects as well as ones at $z>2$ (Grazian et al. 2006; Papovich et al. 2006; 
Conselice et al. 2007; Lane et al. 2007). Quadri et al. (2007) also pointed 
out that the fraction of $z<1.8$ DRGs is 15 per cent at $K<21$ and 50 
per cent at $K<19$. Therefore, from the different clustering properties of the
two populations, we can contrast them at $1<z<2$ where most bright DRGs and 
$i-K>4.5$ EROs are located. The different clustering properties in the bin 
indicate that EROs and DRGs may have different characteristics. In fact, 
Conselice et al. (2007) demonstrated that $1<z<2$ DRGs show a broad range in stellar 
mass and that EROs are more massive than DRGs at the 
same redshift (Conselice et al. 2008). These mass differences can 
explain the stronger clustering of EROs compared to bright DRGs, since massive objects are 
expected to be
more clustered. Furthermore, contamination from low redshift DRGs may lead to the
variation of clustering between our samples and $2<z<3$ DRGs seen in figure 7. The 
fraction of $z<1.6$ DRGs at $K>18.8$ magnitude range is $\sim$40 per cent 
in NMBS redshift distribution. This effect is also verified by the weaker clustering 
of $r-K$ EROs in the previous section. This means that when using a simple colour 
criterion it is difficult to avoid a contribution from different types of galaxy or
galaxies over a wide range in redshift.

Our correlation lengths of DRGs are apparently different from 
previous results. If a single power law is applied for our $K<18.8$ DRGs, the 
correlation length is 9.5$\pm$1.0 $h^{-1}$ Mpc. This value is smaller than 
14.1$_{-2.9}^{+4.8}$ $h^{-1}$ Mpc with $\sigma=$0.5 redshift distribution 
or 11.1$_{-2.3}^{+3.8}$ $h^{-1}$ Mpc using a Gaussian redshift distribution with 
$z=1$ and $\sigma=$0.25 in Foucaud et al. (2007) but is within the error range of
both of these estimates.
These apparent differences may be caused 
by differences in the redshift distribution. The NMBS redshift distribution is
broader and complicated than that assumed in Foucaud et al. and
is likely to better reflect the true DRG redshift distribution.
Quadri et al. (2008) measured $r_{0}=$10.6$\pm$1.6 
$h^{-1}$ Mpc on large scale for 2$<z<$3 DRGs. This is consistent with our results 
for faint DRGs, but smaller than for our brighter DRGs. The improvement
in photometric redshift measurement for galaxies at  $z>1$  that the NMBS provides is considerable.
Future broad band studies will benefit from the NMBS constraints on
redshift distributions.

\subsection{Cosmic variance}

% figure : distribution of EROs
\begin{figure}
\includegraphics[width=7.5cm]{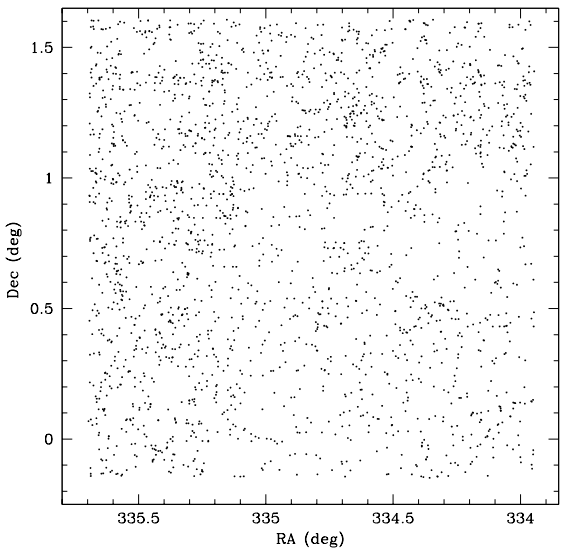}
\caption{The spatial distribution of $K<18.8$ and $i-K>4.5$ EROs.}
\end{figure}

% figure : clustering of EROs depended on field sizes
\begin{figure*}
\includegraphics[width=8.5cm]{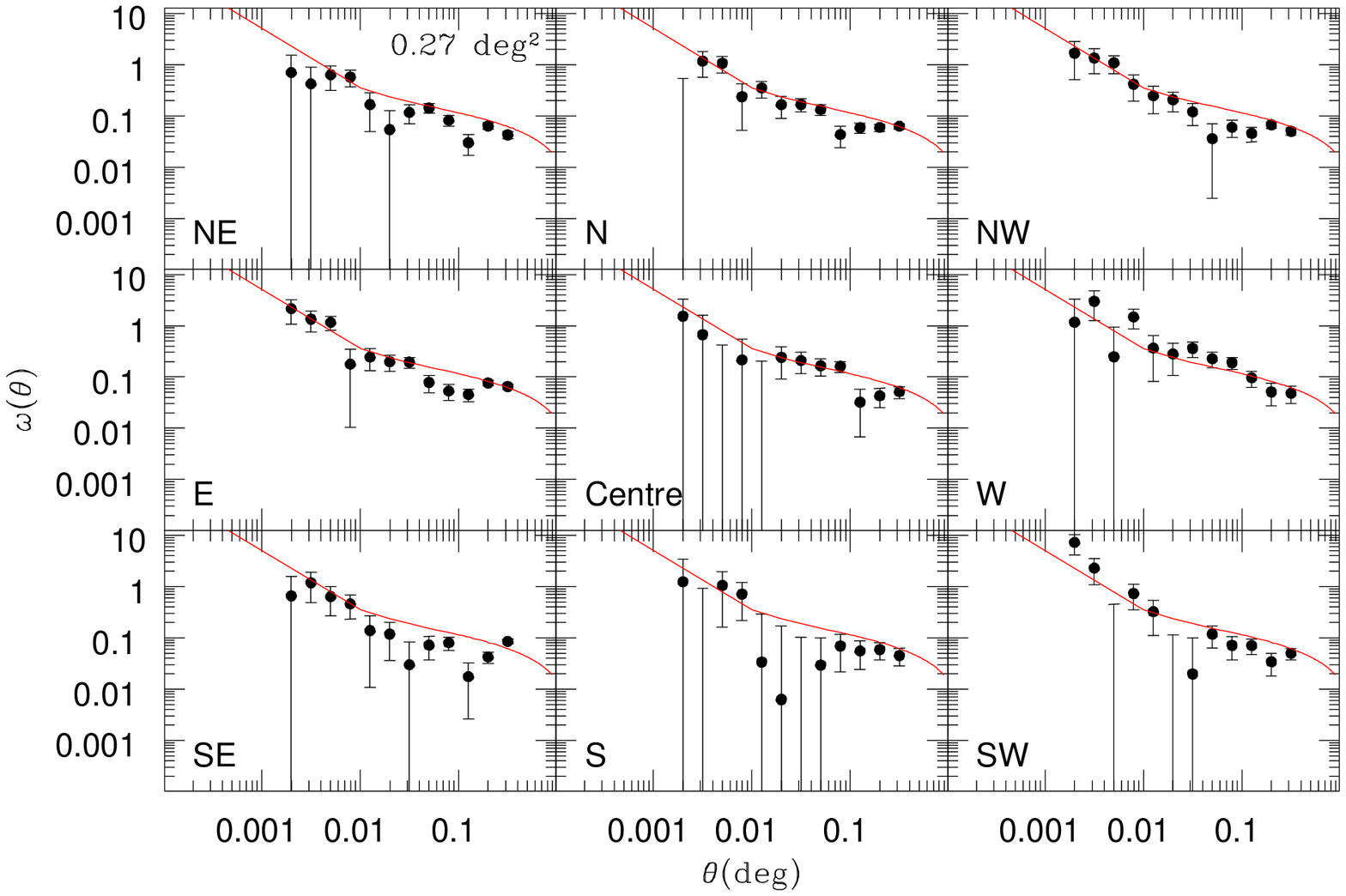}
\includegraphics[width=8.5cm]{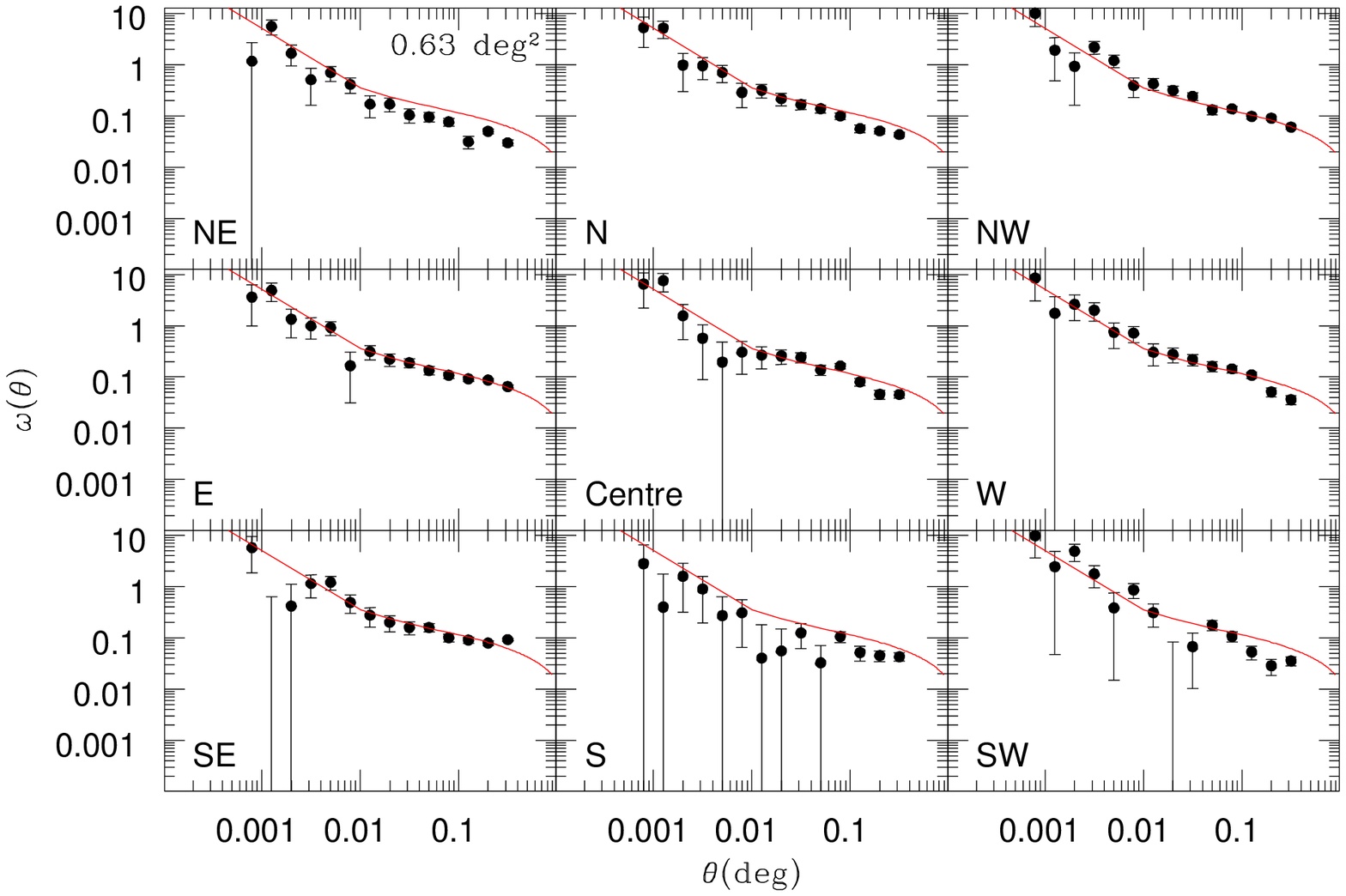}
\caption{The angular correlation functions corrected for the integral constraint of EROs for 0.27 
deg$^{2}$ (left) and 0.63 deg$^{2}$ (right) fields at each position. The solid lines 
indicate the best fit of $K<18.8$ and $i-K>4.5$ EROs in the whole field.}
\end{figure*}

Cosmic variance is an important source of systematic error
in the investigation of the 
high redshift universe (Somerville et al. 2004). In particular, it is a significant 
contribution to the uncertainty in galaxy number counts and luminosity function 
(Somerville et al. 2004; Trenti \& Stiavelli 2008). Somerville et al. (2004) 
defined the relative cosmic variance with mean and variance of number counts.

\begin{equation}
\sigma_{v}^2 \equiv \frac{\langle N^2 \rangle - \langle N \rangle^2}{\langle
N \rangle^2} - \frac{1}{\langle N \rangle}
\end{equation}

The last term gives the correction for Poisson shot noise. Garilli et al. 
(2008) compared cosmic variances defined by the above equation and correlation 
functions from the VIMOS VLT Deep Survey (VVDS) data, and argued that the values 
were in agreement. Therefore, we simply used the above equation to estimate
the cosmic variance.

We divided the SA22 field into 9 sub-fields using two criteria, CTIO Blanco 
field of view ($\sim$0.36 deg$^{2}$) and WFCAM field of view ($\sim$0.8 
deg$^{2}$), with $i-K>4.5$ and $K<18.8$ EROs. However the actual masked 
average areas were $\sim$0.27 deg$^{2}$ and $\sim$0.63 deg$^{2}$ for Blanco 
and WFCAM size fields, respectively. The number density of 
EROs in each field was calculated and used in equation 8. We note that the 
field-to-field ratios of number density were consistent with those for various 
brighter magnitude limits. 
Therefore, the field-to-field variation at this magnitude range is not 
a systematic effect, unlike a change in the limiting magnitude or area. The measured cosmic 
variances were 0.30 and 0.20 for Blanco and WFCAM field size, respectively.

In fact, it is known that the spatial distribution of EROs is inhomogeneous 
(Daddi et al. 2000; Kong et al. 2006), at least in part due to the strong
clustering that EROs exhibit. 
Figure 16 shows the spatial distribution 
of $i-K>4.5$ and $K<18.8$ EROs. As checked by the number density variation of 
each sub-field and figure 16, the distribution 
of our EROs is also inhomogeneous. There are some overdense regions in the 
northern part of the field. Moreover, figure 17 shows the angular 
correlation functions corrected for the integral constraint of $i-K>4.5$ and $K<18.8$ EROs in each sub-field by 
Blanco size (left) and WFCAM size (right). The solid lines indicate the best 
fit of the correlation 
function for $i-K>4.5$ and $K<18.8$ EROs in the whole field. It appears that the
correlation functions show the most variation on Blanco field size scales.  In 
particular, the standard deviations of the amplitudes of angular correlation 
function on large scales are 0.011 and 0.007 for CTIO and WFCAM size 
sub-fields, respectively. Also, the standard deviations of the 
correlation lengths are 
4.5 and 2.8 $h^{-1}$ Mpc respectively. These results demonstrate that 
cosmic variance for these field sizes can significantly affect the uncertainty 
of the measured clustering strength and is likely to have been the dominant
source of error in previous clustering analyses of high redshift galaxies. 
It is apparent that a large-area survey 
is important not only to confidently measure number counts but also to 
investigate clustering properties.

\section{Conclusions}

We have used near-IR images from UKIDSS DXS DR5 and $gri$ optical images from 
CTIO 4m Blanco telescope to 
investigate the clustering properties of EROs and DRGs in $\sim$ 3.3 deg$^{2}$ 
SA22 field. This is the largest area survey of such galaxies to date, and
using the precise redshift distributions from the NMBS we have made
the most accurate measurements of the cluster of EROs and DRGs.
The results are summarised as follows;

\begin{enumerate}

\item Colour selection criteria were applied to extract EROs and DRGs. 
In total 5,383 EROs with $i-K>4.5$, $i<24.6$ and $K<20.0$ were selected. 
In addition, 
3,414 DRGs were extracted by a $J-K>2.3$ with $J<22$ and $K<19.7$ limits. 
The number density of EROs was well matched to previous studies
once the differences in selection method were taken into account.
Similarly, the number density of DRGs was very well matched with 
the results from the UKIDSS UDS field.

\item Both populations showed strong clustering properties. Those of 
EROs are best described by a double power law with inflection at $\sim$ 
0.6$'$-1.2$'$. Assuming a power law, 
$\omega(\theta)=A_{\omega}\theta^{-\delta}$, ($A_{\omega}$, $\delta$) of 
$K<$ 18.8 and $i-K>$ 4.5 EROs were (0.00414, 0.99) and (0.04205, 0.40) for 
small and large scales respectively. 

\item Additionally a double power law is required to fit the angular 
correlation function of DRGs with $\delta=$1.38 and 0.47 for small and large 
scale respectively.  Our relatively bright magnitude limit
samples are diluted by $1<z<2$ DRGs, 
so our clustering  shows different trends when compared to deeper samples
dominated by $2<z<3$ DRGs.

\item The angular two-point correlation function of EROs shows clear trends with 
different magnitude limits, although those for faint samples may be 
dominant by relatively blue EROs due to the optical limit. With a fixed power 
law slope, the amplitude for 
small scales decreased at fainter magnitudes, but that for large scales was 
invariable with magnitude. These trends were also confirmed by the real space 
correlation length. On the other hand, with variable slopes, the correlation 
function at bright limits is steeper than for samples with fainter limits. 

\item The colour limited correlation function of EROs presents slightly different 
features from the magnitude limited function. With a fixed slope, clustering amplitudes 
and real space correlation lengths for small and large scales were increased 
with redder colours. However, slopes were comparable between various colour cuts.

\item The EROs were classified into OGs and DGs by their $i-K$ vs. $J-K$ colours. The 
correlation functions of magnitude limited OGs show an apparent break at 
0.6$'$-1.2$'$ and similar amplitude at large scales. The redder ones have 
stronger clustering. However, the functions for DGs, 
showed much weaker clustering. The relative proportion of
OGs and DGs with colour and magnitude can explain the 
different trends seen in the clustering of the full sample of EROs.

\item EROs selected either with $r-K$ or $i-K$ colours show different 
correlation functions, especially on large scales. The EROs selected by $i-K>3.96$ are 
more clustered than those by the $r-K$ selection criteria. This may be caused by 
the different redshift distribution, since the $r-K$ criterion extracts more 
low redshift galaxies than the $i-K$ criterion.

\item EROs are more clustered than DRGs over the same redshift range ($1<z<2$).
This is evidence that the two populations at this redshift are different
and EROs are likely to be intrinsically more massive than DRGs.

\item  By dividing the full survey field in to sub-fields of different sizes 
we demonstrate that cosmic variance is a significant issue for measurements of 
correlation function and is likely to have been the
dominant source of error in previous measurements of high redshift red galaxy clustering.

\end{enumerate}

The results from this analysis illustrate the importance of sampling
the widest possible fields in the near-infrared in order to recover 
representative clustering properties of distant galaxies. 
In the near future the combination
of UKIDSS and VISTA surveys will cover more than an order of magnitude larger
area to comparable depth. Our ability to extract the clustering of
EROs in these areas is limited only by the depth of comparable
optical imaging. We will be applying the same analysis to the
other three DXS fields and making use of the additional
{\it Spitzer} data that these fields enjoy to refine the
selection into narrower photometric redshift slices and fitting these
the Halo Occupation Distribution models.

Finally, ERO
samples are now of sufficient size to offer direct tests to
galaxy formation models in terms of number density {\it and} clustering
so future comparisons to semi-analytic simulations will be
more powerful (Gonzalez-Perez, in preparation).

\section*{Acknowledgements}

Authors thank referee for comments improving the presentation and content of the paper. 
We thank the NMBS team for sharing their photometric
redshifts of EROs and DRGs in advance of publication.
This work is based on 
the data from UKIRT Infrared Deep Sky Survey. We are grateful to UKIDSS team, 
the staff in UKIRT, Cambridge Astronomical Survey Unit and Wide Field 
Astronomy Unit in Edinburgh. The United Kingdom Infrared Telescope is run by 
the Joint Astronomy Centre on behalf of the Science and Technology Facilities 
Council of the U.K. Cerro Tololo Inter-American Observatory and National 
Optical Astronomy Observatory, managed by the Association of Universities 
for Research in Astronomy, under contract with the National Science Foundation. 
JWK thanks for support from the Korean Government's Overseas Scholarship. 
We thank U. Sawangwit for providing data.

\label{lastpage}

\end{document}